\documentclass[manuscript]{acmart}

\acmJournal{TACO}

\usepackage{xspace}
\usepackage{amssymb}   

\newcommand{\ckv}{\ensuremath{c^{\mathrm{KV}}}\xspace}
\newcommand{\mla}{MLA\xspace}
\newcommand{\Mq}{\ensuremath{M_q}\xspace}
\newcommand{\dqk}{\ensuremath{d_{qk}}\xspace}
\newcommand{\route}{\textsc{route}\xspace}
\newcommand{\fetch}{\textsc{fetch}\xspace}
\newcommand{\local}{\textsc{local}\xspace}
\newcommand{\mape}{\text{MAPE}\xspace}

\begin{document}

\title[Move the Query, Not the Cache]{Move the Query, Not the Cache: Characterizing Cross-Instance
       Latent and Sparse Attention Redistribution Across GPU Fabrics}

\author{Bole Ma}
\affiliation{%
  \institution{Erlangen National High Performance Computing Center (NHR@FAU)}
  \city{Erlangen}\country{Germany}}
\email{bole.ma@fau.de}

\author{Jan Eitzinger}
\affiliation{%
  \institution{Erlangen National High Performance Computing Center (NHR@FAU)}
  \city{Erlangen}\country{Germany}}
\email{jan.eitzinger@fau.de}

\author{Harald K\"ostler}
\affiliation{%
  \institution{Erlangen National High Performance Computing Center (NHR@FAU)}
  \city{Erlangen}\country{Germany}}
\email{harald.koestler@fau.de}

\author{Gerhard Wellein}
\affiliation{%
  \institution{Erlangen National High Performance Computing Center (NHR@FAU)}
  \city{Erlangen}\country{Germany}}
\email{gerhard.wellein@fau.de}

\renewcommand{\shortauthors}{Ma et al.}

\begin{abstract}
Frontier LLMs increasingly decide what a query attends to with a
sparse-attention \emph{indexer} that picks a few KV-cache blocks per query: attention's unit is now a small, reusable chunk. Agentic workloads
hammer it: many sub-agents query one large codebase, reusing the
same blocks. When that corpus outgrows one GPU it is partitioned across
instances, so a query and the blocks it selects often sit on \emph{different}
GPUs: answering it means attention \emph{across} instances.

The reflex of prior cross-instance KV systems is to \emph{move the cache}: pull
the selected blocks to the requester. Multi-head Latent Attention (\mla) inverts
the arithmetic, compressing each token's key and value into one
narrow vector, so a routed query row is only $\approx$1\,KB,
smaller than the chunk it attends; routing the query is then often cheaper than
moving the cache. Which primitive wins, over which
fabric and request shape, is uncharted, least of all on
\emph{device-initiated} RDMA (IBGDA) that makes per-request cross-node transfers
cheap.

We characterize cross-instance \mla attention on a real multi-node H100
cluster, distilling two reusable artifacts: a
topology-aware cost model (probe / transfer / compute / return / merge) and a closed-form
\route/\fetch/\local{} predicate, whose constants we measure on real
IBGDA, where the model tracks batched round-trips to within $\sim$7\%. At
decode it routes the query, trading the cost of moving the cache (a $\approx$3\,ms re-adaptation splice for a
contiguous chunk, or a scattered gather under selection) for a tens-of-microsecond round trip, and picks the fabric by probe latency, not peak
bandwidth. We instantiate the cost model and predicate for \mla, but neither is
\mla-specific: they apply wherever compression or sparse selection shrinks
attention to small chunks (DeepSeek-V3.2, V4, and GLM-5.1 today). Extending
them to a new architecture requires measuring just two coefficients: the
routed payload and \fetch's move-the-cache cost.
\end{abstract}



\maketitle

\section{Introduction}\label{sec:intro}
Modern LLM serving organizes KV state around \emph{prefix reuse}: a
tenant's conversation builds a prefix tree, KV is cached per token, and
reuse happens within one instance, one tenant, one
conversation~\cite{vllm,sglang}. A different reuse pattern is emerging
one level up, at the \emph{provider-curated canonical content} layer. A
provider that pre-prefills a canonical corpus (a body of case law,
public companies' annual and quarterly reports, the trending code
repositories, or a frozen documentation snapshot) into Multi-head Latent
Attention (\mla) compressed-cache (\ckv) form can serve any tenant whose
prompt touches that content from the precomputed cache rather than
re-prefilling it. The unit of reuse is no longer one conversation's
prefix but a canonical chunk shared across tenants and requests. Local prefix
caching cannot capture this reuse: each request precedes the shared chunk with
\emph{different} content, so a contiguous prefix match breaks before reaching
it, and the chunk is reusable only through cross-instance discovery by canonical
id, not by extending a local cache.

This regime breaks a sizing assumption. A few thousand canonical chunks
fit in one accelerator's HBM, but the hot tail grows (the actively-served
slice of these corpora runs to hundreds of gigabytes of \ckv, the full
corpora orders of magnitude more) and the chunks are read by tenants spread
across many serving instances.
Once the canonical store exceeds a single instance's HBM it \emph{must}
be partitioned across instances, and answering a request then routinely
requires attending over \ckv that lives on \emph{another} instance.
The same partitioning arises one scope down: a single tenant's private
corpus (an enterprise knowledge base, a large codebase, or an M\&A
due-diligence data room) can outgrow one instance too, reused across that
tenant's requests rather than across tenants. The agentic workload sharpens
this to a single artifact: one large private document (a codebase, a contract
set, a long report) pinned once as an immutable prefix and queried by many
concurrent sub-agents, each forking it copy-on-write and appending only its own
short suffix. The shared \ckv{} then dwarfs any one agent's appended tokens and
is attended by all of them at once: the byte asymmetry above, here also a
fan-in.
Cross-instance attention becomes a steady-state operation, not an
exception.

How should it be done? The reflex (taken by essentially every
cross-instance KV system to date~\cite{mooncake}) is to \emph{move
the cache}: pull the remote chunk's \ckv over the network, splice it
into the local cache, and attend locally. This reflex carries a habit over from dense attention: bring the cache to
wherever the model is already running, which pays off when the moved cache is
reused by a long local decode. \mla quietly changes the
arithmetic for \emph{per-request} cross-instance attention. Its defining feature is a narrow latent: each
token's key and value collapse into one compressed vector
(the same $\dqk{=}576$ in DeepSeek-V2-Lite and the frontier V3/R1), so a single query row is only
$\approx$1\,KB on the wire while the chunk it attends to holds thousands
of equally-wide cache vectors, three orders of magnitude more data. \mla is,
moreover, a frontier default rather than a niche choice: DeepSeek (V2 through the
sparse-attention V3.2)~\cite{deepseekv2,dsa}, Kimi~K2.6~\cite{kimik2,kimik26}, and the GLM
family (from GLM-4.7-Flash onward, with GLM-5.1 pairing \mla with sparse
attention)~\cite{glm5,glm51} all build on \mla or a close latent-attention variant, so
the byte asymmetry we exploit is broadly shared. Because our cost model depends
on the model only through the wire payload $q{+}p$ (a function of $\dqk$ and
$d_v$), it instantiates to any of them from published dimensions, with
DeepSeek-V2-Lite as our measured instance. When
the query is the
small object, the textbook move inverts: it is cheaper to \emph{route
the query} to the instance holding the \ckv, compute partial attention
there, and merge the small partial back. The routing primitive itself
is not new; it was introduced for sharded KV and since extended to
\mla{} and to single-job context parallelism~\cite{distattention,helix,cp_meta}
(\S\ref{sec:related}), but its \emph{economic case under \mla, across
serving instances, on commodity datacenter fabric} has not been established.

That case is not obvious, for two reasons. First, routing a query batch
is only cheap if a per-request, cross-node transfer is itself cheap,
which points to \emph{device-initiated} RDMA (NVSHMEM IBGDA), where the
GPU issues the transfer directly with no host round trip. Yet recent
transport work finds NVSHMEM IBGDA can be \emph{slower} than a
host-mediated path for the tiny messages of MoE all-to-all --- behind a
CPU proxy in NCCL~GIN's measurements, with portable engines falling back
to a host proxy where the NIC lacks IBGDA~\cite{ncclgin,transferengine}
(quantified in \S\ref{sec:bg-ibgda}) --- but that work targets MoE
dispatch, not attention. Second, picking
wrong is expensive in \emph{both} directions: move the cache and you can
pay more than re-prefilling from scratch (\S\ref{sec:bg}); route the
query at the wrong batch size and a fixed per-request overhead erases the
wire-byte win (\S\ref{sec:select}). Whether device-initiated RDMA pays
off for cross-instance attention, at what request shape, over which
fabric, and with what cost structure, is uncharted.

This paper characterizes cross-instance \mla attention redistribution on
real hardware (a multi-node H100 cluster over NDR-200\,G IBGDA, spanning
same-leaf and spine-traversing node pairs)
and turns the characterization into a decision rule a serving system
can apply per request. We scope deliberately to the \emph{transport}
question (which primitive, which fabric, at which shape); the serving
system that consumes the rule is a separate concern. Although we measure
\mla, the model's structure and predicate are not \mla-specific:
\emph{compression} (\mla's head-axis latent, or DeepSeek-V4's token-axis
CSA compressor) and \emph{selection} (sparse-attention indexers) are
independent levers that shrink the routable unit, and the predicate is
driven by the resulting byte budget, not the attention variant
(\S\ref{sec:select-general}); under selection, routing the query is the
distributed form of the indexer's own choice. Our central result
is deliberately a \emph{crossover} rather than a universal winner:
routing the query moves $\ge$76\% fewer wire bytes at decode-typical
batches ($\Mq{\le}256$) and avoids the move-the-cache splice outright, yet below a measurable query-batch size a
\emph{splice-free} bytes-back fetch still wins end-to-end --- not on wire
bytes, but because a fixed per-request \emph{host} overhead in our
prototype dominates there, a gap three named transport reductions close
(\S\ref{sec:overhead}). Reporting precisely \emph{where} each primitive
wins is the contribution.

\paragraph{Contributions.}
\begin{itemize}
  \item A \emph{topology-aware redistribution cost model} (probe /
  transfer / compute / return / merge) for cross-instance attention, fit to
  real H100 NDR-200\,G IBGDA on the \mla payload at $\approx$7\% \mape for batched dispatch
  ($\Mq\ge512$) (\S\ref{sec:model}).
  \item The \emph{device-initiated-RDMA (IBGDA) regime} for cross-instance
  attention (as opposed to the MoE-dispatch traffic prior IBGDA work covers) has
  not, to our knowledge, been characterised; we do so on
  H100 SXM5 + NDR-200\,G: the IBGDA-vs-proxy crossover
  (proxy $+40\%$ per-fetch p50 at attention's $\sim$2\,KB payload), the holder-side $K$-stream staging elbow
  ($K{=}8$), and the route-holder's matching \emph{compute}-capacity elbow
  ($N{\approx}8$ requesters, measured with DeepSeek's production \mla{} kernel)
  (\S\ref{sec:regime}).
  \item A closed-form, topology-aware \route/\fetch/\local{}
  \emph{primitive-selection predicate} with measured coefficients, and
  the finding that the selection flips with request shape and
  host-overhead regime: although routing wins the wire-byte comparison
  outright, at our prototype's host overhead a splice-\emph{free} bytes-back
  transfer still wins below an \Mq{} threshold until three named transport
  reductions close the gap (\S\ref{sec:select}).
  \item The \emph{byte-asymmetry} framing (that \mla's narrow
  latent makes query-routing fine-grained-\emph{viable}, not merely
  bandwidth-favorable), which we argue is the property that motivates
  device-initiated RDMA for attention (\S\ref{sec:bg}).
  \item \emph{The selection regime as distributed attention}: where a
  sparse-attention indexer shrinks each query to a few scattered KV blocks
  (DeepSeek-V3.2/V4, GLM-5.1), \route{} \emph{is} that selection made distributed
  (attended in place, no cache re-rotation), verified exact to bf16 noise on
  the production sparse kernel, its holder cost set by the selection budget not
  the store's size (\S\ref{sec:setup-correct},~\S\ref{sec:holder-compute},~\S\ref{sec:select-general}).
  \item A \emph{cross-fabric} measurement: the identical routed dispatch on five
  fabrics (PCIe~Gen4/Gen5, NVLink~3.0 and~4.0, cross-node IBGDA) costs
  $\approx$31--48\,$\mu$s at the decode point. The five single-bottleneck
  fabrics cluster within $1.5\times$ (Figure~\ref{fig:bwsens}) because route-RT
  tracks \emph{single-block dispatch}, not link peak: a $900$\,GB/s NVLink~4.0
  exercises only $\approx$21\,GB/s of it, no faster than a PCIe link, so at
  decode it barely leads the cross-node IBGDA path. In the multi-holder gather of a
  scattered selection, each cross-socket-PCIe holder costs \fetch{} $\approx$36\%
  more (a UPI wire penalty the non-blocking fabrics avoid) while \route{} ships
  one query regardless, so routing's edge widens where the fabric is weakest
  (Figure~\ref{fig:select-regime}(a)).
\end{itemize}

\section{Background and Motivation}\label{sec:bg}

\subsection{Cross-instance attention: route, fetch, or recompute}
\label{sec:bg-transport}
When a decoding query must attend to a context chunk residing on another
instance---because prefill and decode are disaggregated, because a hot prefix
is shared across sessions, or because the cache is pooled across the
cluster---the system faces a per-chunk transport choice. It can \fetch{} the
chunk to the querying instance and attend there; \route{} the query to the
chunk's host, compute partial attention there, and return a small partial; or
\local{}, recomputing the chunk from a cheaper form. The three differ only in
what crosses the fabric: the whole chunk, a query row plus its partial, or
nothing.

Prior cross-instance work resolves this in setting-specific ways, none of which
carries over to compressed or selective attention at decode. Prefill/decode
disaggregation moves the cache, but as a \emph{one-time handoff}: the decode
worker takes ownership of the sequence, so the transfer amortises over the
whole generation and overlaps with compute on fast links---cheap by
amortisation, not by per-step byte economy~\cite{mooncake,distserve}. Where
contexts are long and links slow, Infinite-LLM's DistAttention instead routes
the query for \emph{dense, contiguous} attention, shipping a query row plus
online-softmax statistics rather than the cache, and reporting far lower
per-step communication than transferring the cache each step on standard,
non-\mla{} models~\cite{distattention}. A third, very recent line
targets \emph{selective} attention but stays \emph{within a node}: because
top-$k$ selection leaves the KV working set fragmented and hard to
prefetch~\cite{dsaccess}, systems such as Fluxion and FlexiCache coordinate the
scattered gather between GPU and CPU-resident cache over
PCIe~\cite{fluxion,flexicache}, the latter leaving the distributed case to
future work.

Our work turns on a structural point: compression and selection both shrink
attention to \emph{small chunks}, which asymmetrically favours \route{}. The
routed payload is small and \emph{fixed}---a query row plus a width-$d_v$
partial---independent of chunk size, selection budget $k$, and scatter pattern.
The \fetch{} payload is the opposite: it scales with the chunk and, under
selection, becomes a gather of $k$ non-contiguous entries that must cross the
fabric with poor locality and per-entry metadata. \route{} keeps that
irregular gather \emph{local} to the chunk's host and sends only the regular
query/partial across the fabric, turning a bandwidth- and locality-bound
transfer into a latency-bound round trip---so the choice tracks \emph{probe
latency, not peak bandwidth}, per chunk rather than by fixed policy. \mla{} is
the cleanest instantiation: a routed query row and a cached \ckv{} are the same
\dqk-wide object, making the routed payload minimal and precision-invariant,
and \ckv{} position-invariance is what lets a chunk be reused across sessions
at all. But the choice is not \mla-specific---it recurs wherever an
architecture compresses or selects attention into small chunks (\mla, DSA, and
the CSA/HCA hybrids of DeepSeek-V3.2/V4 and GLM-5.1~\cite{dsa,deepseekv4,glm51}),
which is why we cast the contribution as an architecture-general cost model and
\route/\fetch/\local{} predicate (\S\ref{sec:model}).

\subsection{Three redistribution primitives and their cost shapes}
\label{sec:bg-prim}
\begin{figure*}[t]
\centering
\begin{minipage}[b]{0.58\linewidth}
\centering
\includegraphics[width=\linewidth]{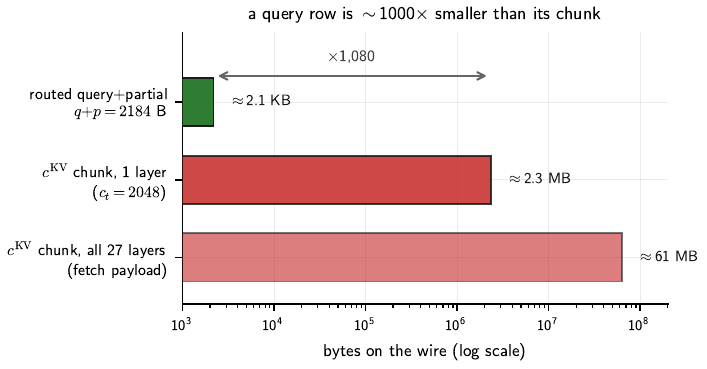}\\
{\footnotesize (a) wire payload: a routed query row vs.\ its \ckv{} chunk}
\end{minipage}\hfill
\begin{minipage}[b]{0.40\linewidth}
\centering
\includegraphics[width=\linewidth]{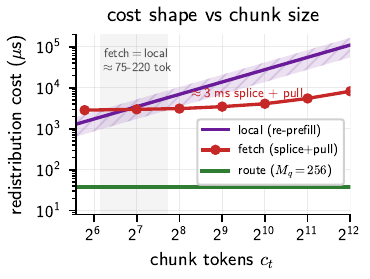}\\
{\footnotesize (b) cost \emph{shape} of the three primitives vs.\ chunk size}
\end{minipage}
\Description{Left (a): horizontal bars on a log axis showing a routed
query+partial (about 2 KB) is roughly a thousand times smaller than a one-layer
(2.36 MB) or all-27-layer (64 MB) c-KV chunk. Right (b): log--log plot of
redistribution cost versus chunk size for three primitives; route is a low flat
line, fetch starts near three milliseconds and rises gently with chunk size,
local re-prefill rises more steeply and crosses fetch around 75--220 tokens,
and route stays about two orders of magnitude below across the whole range.}
\caption{\textbf{(a)}~Payload asymmetry, shown for \mla: because a routed query
row and a cached token are the same narrow object, a routed query$+$partial
($q{+}p{=}2184$\,B) is $\sim$1000$\times$ smaller than the one-layer \ckv{}
chunk a \fetch{} would pull --- the cleanest instance of the \emph{fixed}
routed payload versus chunk-scaling \fetch{} that favours \route{}.
\textbf{(b)}~The load-bearing argument is the cost \emph{shape}, not the byte
count: cost of the three primitives versus chunk size (DeepSeek-V2-Lite, H100, $BW{=}70$\,GB/s,
$\Mq{=}256$). \fetch{} (red, \emph{measured} splice $+$ all-layer pull) carries
a flat $\approx$3\,ms position-adaptation splice plus a gentler chunk-scaling
pull; \local{} (purple, re-prefill, band over
$c\,{\in}\,[0.5,1.5]\,\mu$s/token$\cdot$layer) scales with the chunk and
overtakes \fetch{} only above $\approx$75--220 tokens; \route{} (green) pays
neither and stays about two orders of magnitude below. The cost-shape
asymmetry, not wire bytes alone, is the structural case for routing.}
\label{fig:costshape}
\end{figure*}

\S\ref{sec:bg-transport} named the three primitives by \emph{what} crosses
the fabric; they differ just as much in the \emph{shape} of that cost, which
is what makes the choice hard. \fetch{} (the pull reflex) does more than
transfer $A$'s \ckv: it re-rotates the cached positional encoding to $B$'s
offset (when re-homing the chunk to a new contiguous offset) and splices it
into $B$'s paged cache before attending. \route{} returns a partial that $B$
merges via online softmax~\cite{onlinesoftmax,flashattention}, the
cross-instance form of the routing primitive of DistAttention and
Helix~\cite{distattention,helix}; \local{} is a fresh re-prefill. Beyond the
wire transfer (pulling a chunk's \ckv is
$\approx$2.5\,ms, all $L{=}27$ layers bulk-coalesced into one transfer over the
cross-node NDR-200 link at $\approx$25\,GB/s), \fetch{} pays a \emph{splice} on
top: re-rotating the cached \ckv to the requester's position and scattering it
into the local paged pool. We measure this splice at
\textbf{$\approx$2.8--3.1\,ms per chunk} on DeepSeek-V2-Lite (H100), and
find it \emph{essentially independent of chunk size}: it grows only
$\sim$10\% from 55 to 4096 tokens because it is dominated ($\approx$80\%)
by the per-layer $\delta$-rotation kernel, not by token count. \local{}
(re-prefill) has the opposite shape: its cost scales with the chunk, into
the tens of milliseconds for a 2\,k-token chunk. So \fetch{} carries a
\emph{flat $\approx$3\,ms position-adaptation tax} and \local{} a
\emph{size-scaling recompute}, while \route{} pays \emph{neither}: the
holder's \ckv is already resident at its canonical position, so there is
no splice and no re-prefill. That cost-shape asymmetry, not the wire
bytes alone, is the structural case for routing (Figure~\ref{fig:costshape}).
The adaptation tax is not an
\mla quirk: position-independent caching on standard GQA/MHA models pays
the analogous cost as a small \emph{carved}-prefix recompute per
chunk~\cite{epic,mepic,cacheblend}; on \mla it takes the form of a purely positional \ckv $\delta$-rotation. Both this
$\delta$-rotation and EPIC's sink-repair carve are costs of \emph{contiguous}
reuse: under sparse selection (NSA, DSA)~\cite{nsa,dsa} the chosen entries are
attended at their \emph{canonical} positions, so neither applies and \fetch's only
residual cost is the scattered gather
(\S\ref{sec:setup-correct},~\S\ref{sec:select-general}).

\subsection{Device-initiated RDMA: the uncharted regime}
\label{sec:bg-ibgda}
Systems analyses of post-\mla inference now point to the
\emph{interconnect}, not attention compute, as the emerging bottleneck:
\mla's arithmetic intensity runs about two orders of magnitude above
MHA, pushing on-device attention into a compute-bound regime and
relocating pressure onto the fabric and MoE expert
balancing~\cite{newbottleneck}. That analysis targets MoE all-to-all
dispatch; cross-instance attention redistribution is a second
fabric-bound workload it leaves open, and \route{} bets on exactly
this fabric.
\route{} issues many small, per-request transfers, so its viability
hinges on how cheaply a cross-node transfer can be launched. Two
substrates exist. A \emph{CPU-proxy} path has a host thread fill the
NIC's work-queue entries on the GPU's behalf; \emph{device-initiated}
RDMA (NVSHMEM IBGDA) lets a GPU thread post the transfer itself, taking
the GPU$\to$host$\to$NIC handoff off the critical path. Intuitively
IBGDA should win for latency-sensitive per-request traffic, and for
attention it essentially must, since a host round trip per query batch
would dominate the few-microsecond wire time.

The literature complicates this. For the tiny messages of MoE
all-to-all dispatch, NVSHMEM IBGDA is reported \emph{slower} than
host-mediated paths: NCCL~GIN measures a $24.3\,\mu$s IBGDA round trip
against $18.0\,\mu$s for a CPU proxy and $16.7\,\mu$s for its own
device-initiated GDAKI backend~\cite{ncclgin}, and portable engines fall
back to a host proxy where IBGDA is unavailable on the
NIC~\cite{transferengine}. Two caveats reopen the question for
attention. First, those results are for MoE dispatch, where messages are
tens to hundreds of bytes; an \mla query batch is several kilobytes,
where the fixed per-message issue cost amortizes differently. Second,
the slow path is \emph{NVSHMEM IBGDA specifically}: GDAKI, also
device-initiated, is competitive, so the outcome is empirical, not
settled by first principles. Whether IBGDA's per-request model pays off
for attention-shaped payloads on commodity NDR-200\,G fabric, and how it
composes with holder-side staging, has not been measured.
\S\ref{sec:regime} closes this gap and \S\ref{sec:model} models the cost
so the \route/\fetch/\local{} decision can be made in closed form.

\section{Experimental Setup}\label{sec:setup}

\subsection{Platform and microbenchmark harness}
\label{sec:setup-platform}
Our primary platform is a production cluster of 4$\times$H100 SXM5 nodes,
with direct all-to-all NVLink~4.0 intra-node (six bonded links
per GPU pair, \texttt{NV6}; the 4-GPU HGX board carries no NVSwitch) and InfiniBand (NDR-200)
cross-node, running NVSHMEM~26.3 with IBGDA enabled. A cross-instance run is a
2-node $\times$ 4-PE NVSHMEM job (PE\,0 and PE\,4 the cross-node pair);
because the cluster is a multi-leaf fat-tree we place this pair both
\emph{within} a leaf switch and \emph{across the spine} and measure the
route probe and round trip identical either way (\S\ref{sec:sens-scale}),
so the slice stands in for any cross-NVL link in the fabric rather than a
two-adjacent-node special case. The
device-initiated primitive is block-scope
\texttt{nvshmemx\_putmem\_signal\_nbi\_block}, a one-way write with a
piggy-backed signal; every latency is per-iteration wall clock from CUDA events
over 200 timed iterations after 50 warm-up. The cost model, splice, congestion,
topology, and staging results are all on H100; for the fabric-robustness study
(Figure~\ref{fig:bwsens}) we run the \emph{identical} primitive on three further
GPU types in a separate single-node testbed (A40, PCIe~Gen4 cross-socket; A100,
NVLink/NVSwitch; RTX~Pro~6000, PCIe~Gen5) as intra-node P2P bandwidth anchors,
the H100 cluster supplying the cross-node IBGDA point. (Those partitions are
single-node, so cross-node IBGDA stays the H100 measurement.)

\paragraph{Configuration caveat.} By site policy this cluster runs the
\emph{legacy} (closed) NVIDIA kernel driver: the open GPU kernel module
(\texttt{kmod-nvidia-open-dkms}) destabilised the nodes, and the
GPUDirect Storage kernel module (\texttt{nvidia-fs-dkms}) depends on it,
so both were reverted. Consequently (i)~GPUDirect Storage is unavailable
system-wide, and (ii)~IBGDA runs over the legacy GPUDirect-RDMA path.
Our \emph{absolute} latencies and bandwidth are therefore a
configuration-specific, likely-conservative operating point, not the
fabric's ceiling. This shapes what we claim: the results our conclusions rest on are
the \emph{cost model} (\S\ref{sec:model}) and the \emph{relative}
\route/\fetch/\local{} decision (\S\ref{sec:select}), both invariant to
the absolute fabric constants; absolute numbers are reported as a
calibrated instance and contextualised against published same-class
measurements. A faster driver would lower $T_{\mathrm{probe}}$ and raise
$BW$ --- shrinking \route's cost and \emph{strengthening}, not weakening,
the case for routing.

\subsection{Wire format and what we measure}
\label{sec:setup-wire}
Under Q-routing the requester ships, per attended chunk, a batch of \Mq
absorbed-\mla query rows --- each \dqk-wide ($d_{qk}{=}576$) in bf16 (2\,B), so
$q{=}576{\times}2{=}1152$\,B/row --- and receives a \emph{partial}: the
holder's attention over its resident subset as one $d_v$-wide
($d_v{=}512$) bf16 output row $o$, plus the running max-logit $m$ and
softmax denominator $\ell$ (fp32, 4\,B) that let the requester merge it
exactly, $p{=}512{\times}2{+}2{\times}4{=}1032$\,B/row. The triple
$(o,m,\ell)$ is the sufficient statistic FlashAttention carries between
tiles~\cite{flashattention,onlinesoftmax}, here carried between
instances. We time two quantities: \texttt{sig\_rt},
a one-byte put-plus-signal round trip (the protocol \emph{probe}); and
\texttt{full\_rt}, the full $\Mq{\cdot}q$ send $+\,\Mq{\cdot}p$ return
with both signals. The $\delta$-rotation that aligns each query to the
holder's offset is applied requester-side before enqueue, so the holder is
position-oblivious (\S\ref{sec:bg-prim}). The query-batch sweep covers
$\Mq\in\{1,4,\dots,4096\}$; a payload sweep additionally varies $(q,p)$
over a $10\times$ span ($900$--$8736$\,B/row) to probe the
payload-dependence of bandwidth (\S\ref{sec:model}).

\subsection{Correctness of the routed primitive}
\label{sec:setup-correct}
Before characterising cost we confirm the routed primitive is
numerically faithful, so the characterisation describes a correct
mechanism rather than a degenerate one. On real cross-node IBGDA, a requester
query routed to a holder, attended against the holder's local \ckv,
returned, and merged via online softmax against the requester's own
partial reproduces single-instance \mla attention over the concatenated
cache to \textbf{max-absolute $0.0014$}, inside the $0.05$ bf16-wire
noise floor by a $36\times$ margin. The merge is bit-identical across the
two implementations it must agree on (serving stack and transport),
verified in unit tests for commutativity and the zero-weight identity.

The same faithfulness holds in the distributed-selection regime of
\S\ref{sec:select-general}, where it matters most. Routing a query to
$M$ holders that each own a disjoint, \emph{scattered} subset of a selected
set (every entry left at its canonical decoupled-RoPE position) and
merging the $M$ partials reproduces single-instance \mla attention over the
whole set to fp32 round-off ($\le\!4\times10^{-7}$ max-absolute, invariant to
$M$ up to $8$ and to how the set is partitioned across holders), and to the
same bf16 floor on the wire (a scattered two-instance route--merge lands at
max-absolute $0.0012$). No position adaptation is applied, and none is
admissible: re-homing the scattered selection to contiguous offsets (the
$\delta$-rotation a contiguous-reuse \fetch{} applies) instead
\emph{diverges} from the reference by $25$--$56\%$, confirming that splice is a
property of contiguous reuse, not of selection. These checks use our reference
attention; replaying the same $M$-way merge on \emph{production} kernels ---
DeepSeek's FlashMLA dense-decode kernel~\cite{flashmla}, FlashInfer's paged
MLA~\cite{flashinfer}, and, most directly, FlashMLA's bf16 \emph{sparse}
kernel~\cite{flashmla} (the selection path DSA~\cite{dsa} deploys, where each
holder runs it over a disjoint subset of the query's top-$k$ selected indices)
--- reproduces each kernel's single full-set call to max-absolute $0.002$ across
$M{\le}8$ and selected sets of $512$--$2048$ (all return the log-sum-exp the
merge consumes). The exactness is thus a property of the online-softmax algebra,
not of our reference, and it holds on the very kernel a sparse-attention server
runs.

\section{A Topology-Aware Redistribution Cost Model}\label{sec:model}

\subsection{Cost decomposition}
We model any cross-instance redistribution as additive terms,
\[
T_{\mathrm{redist}}(F,s,B) = T_{\mathrm{probe}}(F) + T_{\mathrm{transfer}}(F,s,B)
  + T_{\mathrm{compute}} + T_{\mathrm{return}}(F,s,B') + T_{\mathrm{merge}},
\]
where $F$ is the fabric (intra-NVL NVLink, cross-NVL IBGDA, intra-node
PCIe), $s$ the device-initiated scope (thread / warp / block), $B$ the
transfer-unit size and $B'$ the return size, $T_{\mathrm{compute}}$ the
holder's partial-attention over its resident subset, and
$T_{\mathrm{merge}}$ the requester's online-softmax recombination. The decomposition, not
its absolute calibration, is the contribution: it localises
\emph{where} a redistribution's cost lives, and hence which primitive and
fabric win in each regime, in a form robust to the absolute constants of
any one configuration (\S\ref{sec:setup}).

\subsection{Per-primitive instantiation}
For Q-routing the instantiation is
\[
T_{\mathrm{route}}(F,\Mq) = T_{\mathrm{probe}}(F) + \Mq\,(q+p)/BW(F)
  + T_{\mathrm{compute}} + T_{\mathrm{merge}},
\]
with $q,p$ the query and partial row sizes (\S\ref{sec:setup}) and
$BW(F)$ the fabric's effective bidirectional throughput; the first two
terms are the \emph{transport} round trip, $T_{\mathrm{compute}}$ the
holder's partial attention ($15$--$37\,\mu$s at decode,
\S\ref{sec:holder-compute}) and $T_{\mathrm{merge}}$ the online-softmax
merge ($\le\!25\,\mu$s). Our fit below calibrates the transport terms,
which carry the $\Mq$- and fabric-scaling and hence the route/fetch
decision; $T_{\mathrm{compute}}$ and $T_{\mathrm{merge}}$ are bounded,
payload-light constants measured separately (\S\ref{sec:regime}), so the
$T_{\mathrm{route}}$ we report as a $\approx$116\,$\mu$s round trip (at $\Mq{=}1024$) is the
transport term: including them adds a bounded tens of microseconds (at
decode, comparable to the transport round trip itself) yet keeps \route{}
more than an order of magnitude below \fetch's $\approx$3\,ms splice. \fetch{} instead
pays $T_{\mathrm{fetch}} = T_{\mathrm{pull}} + T_{\mathrm{splice}}$, where
$T_{\mathrm{pull}} = c_t\,b_{\mathrm{KV}}/BW(F)$ moves the whole
$c_t$-token chunk ($b_{\mathrm{KV}}$ its per-token \ckv size) and
$T_{\mathrm{splice}}$ is the position-adaptation cost (\S\ref{sec:bg-prim});
under sparse selection $T_{\mathrm{splice}}$ vanishes and $T_{\mathrm{pull}}$
becomes a scattered, multi-holder gather (\S\ref{sec:select-general}).
\local{} is a fresh prefill $T_{\mathrm{prefill}}(c_t)$. The serving layer
applies the predicate of \S\ref{sec:select}, $\arg\min$ over the
three, per request.

\subsection{Fit and validation on real IBGDA}
We calibrate the model on real H100 IBGDA. The probe $T_{\mathrm{probe}}$
is the measured \texttt{sig\_rt}: $\mathbf{\approx\!16\,\mu s}$, and it is
\emph{payload-independent} ($16.2/15.9/16.6/16.1\,\mu$s across the four
payloads of Table~\ref{tab:payload}), as a one-way put-plus-signal must
be. The bandwidth $BW$ is read from the large-\Mq{} slope:
$\mathbf{\approx\!25}$\,\textbf{GB/s} effective (the NDR-200 link rate), and it too is
payload-independent across a $10\times$ span of $(q{+}p)$. This
payload-independence is the empirical content of the linear-in-bytes
term: $T_{\mathrm{route}}-T_{\mathrm{probe}}$ grows with $\Mq(q{+}p)$ at a
constant slope, so the model \emph{measures} its own structure rather
than assuming it.

\begin{table}[t]\centering\small
\caption{IBGDA Q-dispatch across a $10\times$ payload span (real \mla
payload bold; other rows vary the payload size on the same transport ---
\emph{synthetic} marks the sub-\mla{} stand-in). The
probe (\texttt{sig\_rt}, the payload-free signal round
trip) and effective bandwidth (effBW $=\Mq(q{+}p)/(\text{full\_rt}-\text{probe})$,
bytes moved per unit transfer time) are payload-independent, the empirical
basis for the linear-in-bytes cost term.}
\label{tab:payload}
\begin{tabular}{rrrr}
\toprule
$q{+}p$ (B/row) & \texttt{sig\_rt} ($\mu$s) & \texttt{full\_rt}@1024 ($\mu$s) & eff.\ $BW$ (GB/s)\\
\midrule
900 (synthetic)        & 16.2 & 62.8  & 24.6\\
\textbf{2184 (real)}   & \textbf{15.9} & \textbf{115.8} & \textbf{24.7}\\
4368 ($2\times$)       & 16.6 & 207.7 & 24.7\\
8736 ($4\times$)       & 16.1 & 389.1 & 24.7\\
\bottomrule
\end{tabular}
\end{table}

\begin{figure*}[t]
\centering
\begin{minipage}[b]{0.60\linewidth}
\centering
\includegraphics[width=\linewidth]{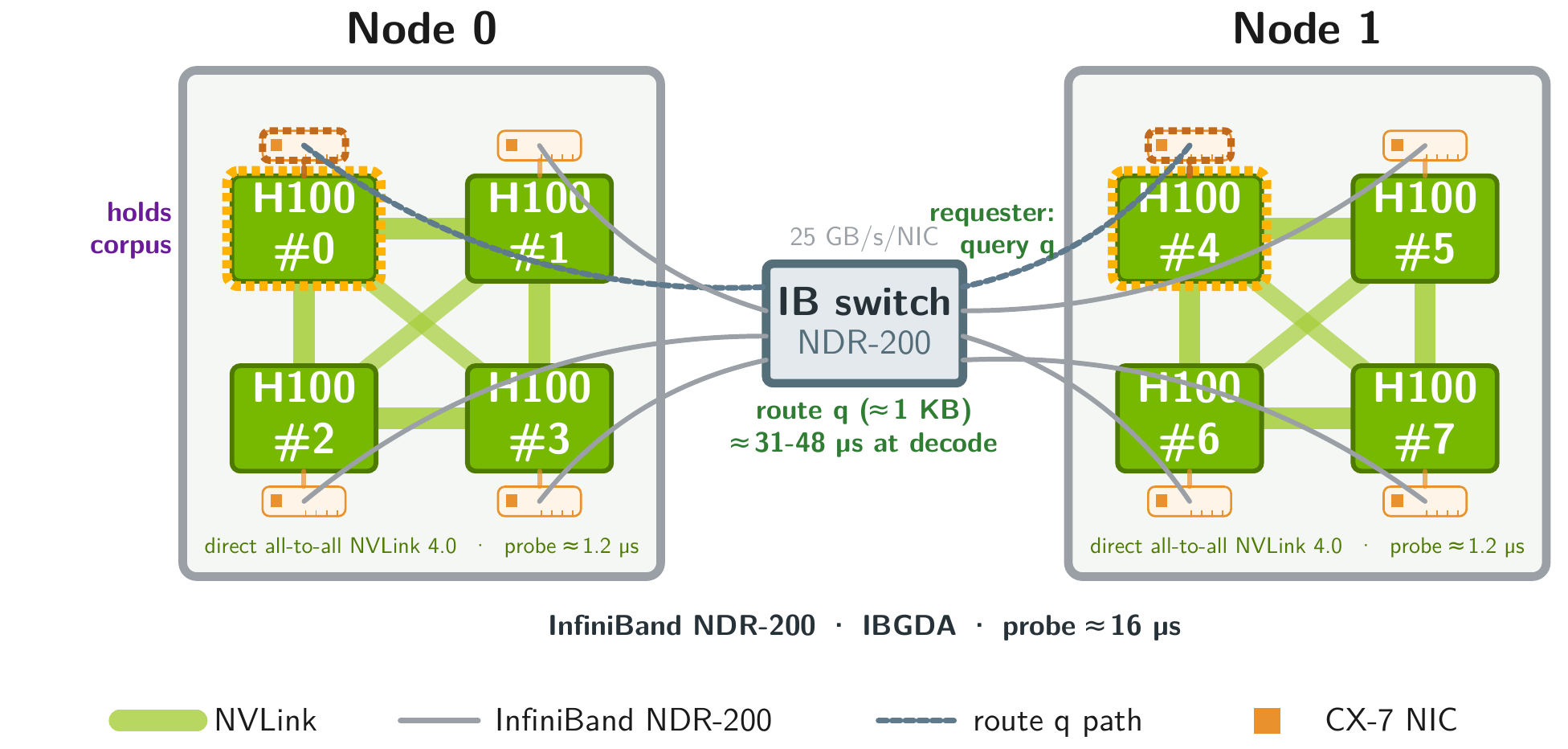}\\
{\footnotesize (a) 2-node $\times$ 4-H100 testbed: direct-NVLink islands,
cross-node IBGDA, the routed query path}
\end{minipage}\hfill
\begin{minipage}[b]{0.38\linewidth}
\centering
\includegraphics[width=\linewidth]{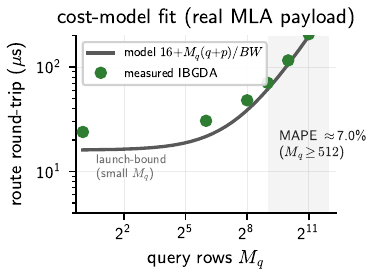}\\
{\footnotesize (b) measured \route{} round trip vs.\ the cost model}
\end{minipage}
\Description{Left (a): schematic of two nodes, each with four H100 GPUs in a
direct all-to-all NVLink mesh and one ConnectX-7 NIC per GPU, joined through a
central InfiniBand switch; a dashed path marks a routed query crossing from a
requester GPU on one node to the corpus-holding GPU on the other. Right (b):
log--log scatter of measured route round-trip versus query rows overlaid with
the cost-model line, matching to about 7\% in the large-batch regime.}
\caption{\textbf{(a)}~The 2-node $\times$ 4-H100 SXM5 testbed: each node is a
direct all-to-all NVLink island (\texttt{NV6} per GPU pair, no NVSwitch) with
one ConnectX-7 NIC per GPU; cross-node traffic is device-initiated IBGDA over
an InfiniBand NDR-200 switch, and the dashed path traces a routed query from a
requester GPU to the corpus holder. The pair is drawn here same-leaf; we
measure it both same-leaf and spine-traversing (cross-leaf) and find the
route cost identical (\S\ref{sec:sens-scale}), so this slice characterises
any cross-NVL link in the fabric. \textbf{(b)}~Cost-model validation: measured
\route{} round trip on real H100 IBGDA versus
$T_{\mathrm{route}}{=}T_{\mathrm{probe}}{+}\Mq(q{+}p)/BW$ with its
\emph{measured} constants ($T_{\mathrm{probe}}{\approx}16\,\mu$s, $BW{\approx}25$\,GB/s;
real \mla payload $q{+}p{=}2184$\,B), tracking the amortised regime
($\Mq\ge512$, shaded) to $\approx$7\% \mape; the small-\Mq{} gap is a fixed
$\sim$9\,$\mu$s kernel turnaround, not a model defect.}
\label{fig:fit}
\end{figure*}

At the real \mla payload ($q{=}1152$, $p{=}1032$), $T_{\mathrm{route}}$ at
$\Mq{=}1024$ is $\mathbf{\approx\!116\,\mu s}$ measured; the model
$\big(16 + \Mq(q{+}p)/BW\big)$, with its \emph{measured} constants and no
refit, reproduces the amortised regime ($\Mq\ge512$) to $\mathbf{\approx\!7\%}$ \mape
($\approx$3\% for $\Mq\ge2048$), the residual a fixed $\sim$9\,$\mu$s kernel
turnaround beyond the probe that the linear term omits, which grows the
small-\Mq{} error (Figure~\ref{fig:fit}). Against the position-adaptation cost of the
equivalent \fetch{} (the $\approx$3\,ms splice of \S\ref{sec:bg-prim}),
routing is $\mathbf{\approx\!26\times}$ cheaper per requester at
$\Mq{=}1024$, rising to $\sim\!125\times$ at $\Mq{=}1$, the single-token decode
step that dominates generation. The practical
import of the few-percent fit is that the selection predicate of
\S\ref{sec:select} can be \emph{evaluated}, not profiled: a scheduler
plugs a fabric's two measured constants into the closed form and obtains a
per-request \route/\fetch/\local{} cost arithmetically, with no online
calibration.

\paragraph{On the absolute level.} These constants are specific to the
legacy-driver configuration of \S\ref{sec:setup}; published IBGDA
characterisations on other H100$+$InfiniBand systems report small-message
round trips ranging from single to tens of microseconds depending on NIC
and driver~\cite{ncclgin}, so our figures sit within the expected band
while plausibly leaving headroom under the open driver. The
\route-vs-\fetch{} decision, however, turns on the \emph{ratio}
$T_{\mathrm{route}}/T_{\mathrm{fetch}}$, not on either absolute, and that ratio
stays overwhelmingly in routing's favour on every fabric we measure: a
decode-sized route costs $\approx$31--48\,$\mu$s across the five
(\S\ref{sec:sens-scale}, Fig.~\ref{fig:bwsens}(b)): a fabric-specific probe
of $1$--$16\,\mu$s plus a dispatch-bound transfer, still more than an order
of magnitude below \fetch's $\approx$3\,ms splice on each. Our numbers are thus
a conservative anchor for the predicate of \S\ref{sec:select}.

\paragraph{One model, many fabrics.} The affine form is not specific to IBGDA:
re-fitting only its two constants reproduces the four other sweeps to comparable
accuracy (Table~\ref{tab:fabricfit}): $\approx$2\% \mape on RTX~Pro~6000
PCIe~Gen5, $\approx$3\% on H100 NVLink~4.0, and $\approx$4\% on A100
NVLink~3.0, bracketing IBGDA's $7\%$. The model is thus fabric-general in
\emph{structure}, its two per-fabric constants (Table~\ref{tab:fabricfit})
splitting along orthogonal axes: $T_{\mathrm{probe}}$ tracks fabric
\emph{latency}, while $BW$ is the single-block \emph{dispatch} rate, far below
the link peak on the fast ones because one put$+$signal block cannot saturate a
wide wire. The one departure is the A40 PCIe~Gen4 large-$\Mq$ tail: its
same-socket fit goes super-linear past $\Mq{=}2048$, and an isolated cross-socket
flow is slow but non-reproducible, an idle-path warm-up artifact that vanishes
under load (\S\ref{sec:sens-scale}), not a steady bandwidth limit.

\begin{table}[t]\centering\small
\caption{The affine cost model
$T_{\mathrm{route}}{=}T_{\mathrm{probe}}{+}\Mq(q{+}p)/BW$ re-fits all five
measured fabrics with its own two constants, to $\approx$2--7\% \mape in the
amortised regime. The two constants split cleanly: $T_{\mathrm{probe}}$ is
\emph{fabric-specific} (NVLink $\sim$1\,$\mu$s, PCIe a few, cross-node IBGDA
$16$\,$\mu$s), while $BW$ is the single-block \emph{dispatch} rate
($\sim$18--25\,GB/s on every fabric we tested), far below the link peak on the
fast ones. The A40 PCIe~Gen4 path is the edge case: same-socket it is
wire-bound and fits to $\approx$3\% through $\Mq{=}2048$ before a super-linear
tail at $4096$ (a second queueing bottleneck) lifts the amortised \mape to
$13\%$. (An \emph{isolated} cross-socket A40 flow is slower and
non-reproducible, but equals the same-socket cost under any load: an
idle-path warm-up artifact, not a steady limit (\S\ref{sec:sens-scale}).)}
\label{tab:fabricfit}
\begin{tabular}{lrrrr}
\toprule
fabric & $T_{\mathrm{probe}}$ & $BW$ & \mape & \mape\\
       & ($\mu$s) & (GB/s) & ($\Mq{\ge}512$) & (full)\\
\midrule
H100 IBGDA (cross-node)        & 16  & 25 & 7\%  & 16\%\\
H100 NVLink~4.0 (intra-node, NV6 direct) & 1.2 & 21 & 3\%  & 18\%\\
A100 NVLink~3.0 (NVSwitch)     & 1.6 & 18 & 4\%  & 22\%\\
RTX~Pro~6000 PCIe~Gen5         & 4.8 & 22 & 2\%  & 7\%\\
A40 PCIe~Gen4 (same-socket)    & 8.7 & 19 & 13\% & 16\%\\
\bottomrule
\end{tabular}
\end{table}

\section{Primitive Selection: \texorpdfstring{\route}{ROUTE} vs
         \texorpdfstring{\fetch}{FETCH} vs \texorpdfstring{\local}{LOCAL}}
\label{sec:select}

\subsection{The closed-form predicate}
The serving layer chooses per (chunk, request) by evaluating the three
costs of \S\ref{sec:model} and taking the minimum. With the constants we
measure --- $T_{\mathrm{probe}}{\approx}16\,\mu$s and $BW{\approx}25$\,GB/s for
\route{}; a flat $\approx$3\,ms splice plus a $\approx$2.5\,ms all-layer
pull for \fetch{} (\S\ref{sec:bg-prim}), or a scattered gather under selection
(\S\ref{sec:select-general}); and a re-prefill of
$\approx c_t\,L\,c$ for \local{} ($L{=}27$ layers,
$c\!\approx\!0.5$--$1.5\,\mu$s per token$\cdot$layer) --- the predicate
takes a simple shape. \fetch{} overtakes \local{} only above a small
chunk ($c_t \gtrsim 75$--$220$ tokens, where the flat 3\,ms splice
undercuts re-prefill), but \route{} undercuts \emph{both} across the
whole range: at a 2\,k-token chunk its $\approx$116\,$\mu$s wire cost
($\Mq{=}1024$) is more than an order of magnitude below \fetch's $\approx$3\,ms
and over two orders below \local's tens of milliseconds. \route{} cedes only in the
corner where it would ship more than the chunk itself: a query batch
larger than the chunk it attends ($\Mq \gtrsim c_t$, \S\ref{sec:bg-transport}),
where \local{} or \fetch{} wins. The predicate is closed-form and
evaluated in microseconds; the rest of the paper characterises its
inputs.

\subsection{The \texorpdfstring{\Mq}{Mq}~\texorpdfstring{$\times$}{x}~chunk-tokens crossover map}
\begin{figure*}[t]
\centering
\begin{minipage}[b]{0.57\linewidth}
\centering
\includegraphics[width=\linewidth]{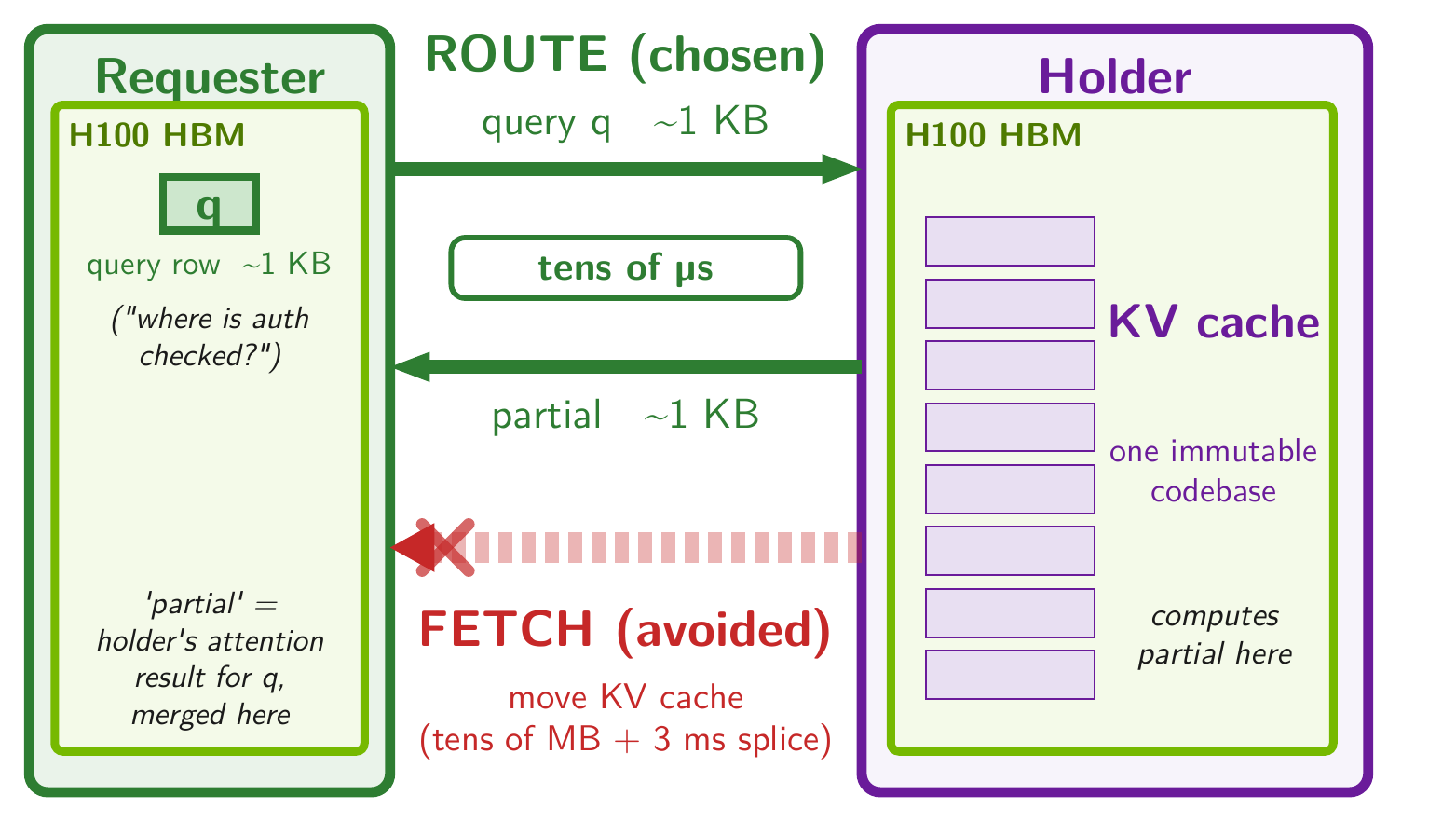}\\
{\footnotesize (a) route the query, not the cache}
\end{minipage}\hfill
\begin{minipage}[b]{0.41\linewidth}
\centering
\includegraphics[width=\linewidth]{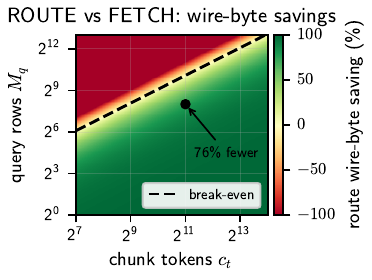}\\
{\footnotesize (b) wire-byte winner over $(\Mq, c_t)$}
\end{minipage}
\Description{Left (a): a requester GPU and a holder GPU, each shown with its H100
HBM; the holder's HBM is packed with the c-KV cache and the requester's holds one
tiny query row. Route ships the small query and a small partial back (chosen);
fetch would move the whole multi-megabyte cache (avoided). Right (b): heatmap
over query rows versus chunk tokens, green where routing moves fewer wire bytes
and red where fetching is cheaper, with a dashed break-even diagonal; a decode
step on a 2k-token chunk at 256 query rows sits deep in green at about 76\%
fewer routed bytes.}
\caption{\textbf{(a)}~The choice, with both operands resident in H100 HBM: a
holder owns a large \ckv{} corpus, a requester has a $\approx$1\,KB query row;
\route{} moves the query and merges a small partial back (chosen), while
\fetch{} would move the whole multi-megabyte cache (avoided).
\textbf{(b)}~\textsc{route} vs \textsc{fetch} on \emph{wire bytes} over the
$(\Mq, c_t)$ grid (DeepSeek-V2-Lite, bf16): green where routing moves fewer
bytes, red where pulling the chunk does. The dashed break-even line is
$\Mq = c_t\,b_{\mathrm{KV}}/(q{+}p)$; a decode step attending a hot 2\,k-token
chunk ($\Mq{=}256$) sits at $76\%$ fewer routed bytes.}
\label{fig:crossover}
\end{figure*}

The byte-level view sharpens where \route{} and \fetch{} cross. \route{}
moves $\Mq(q{+}p)$ bytes while pulling the chunk moves
$c_t\,b_{\mathrm{KV}}$ (\S\ref{sec:bg-transport}), so on wire bytes the winner
flips at $\Mq = c_t\,b_{\mathrm{KV}}/(q{+}p)$ (Figure~\ref{fig:crossover}).
At $c_t{=}2048$, routing
moves $\ge$76\% fewer wire bytes for $\Mq\le256$ (76\% at $\Mq{=}256$, rising
toward $\sim$100\% for smaller batches), and breaks even near
$\Mq\approx10^3$. A serving layer attending a hot 2\,k-token chunk with
the $\Mq\le256$ of a decode step (every token a requester generates against that
cached chunk is one such step) therefore sits deep in \route{}'s region
on wire bytes --- and, once \S\ref{sec:select}'s splice and re-prefill
are added, by a wider margin still on total cost.

\subsection{When host overhead, not wire bytes, decides}
\label{sec:overhead}
Wire bytes are not end-to-end latency. Measured through our (Python)
client on real IBGDA, \route's time-to-first-token scales as
$\approx 3.5\,\text{ms} + 12.5\,\mu\text{s}\cdot\Mq$: a fixed \emph{host}
overhead dwarfs the microsecond-scale wire cost. Against the
\emph{transport-level} fetch primitive (a plain bytes-back transfer
with a single response), \route's three-put response $(o,m,\ell)$ plus
holder-side attention (a bounded $15$--$37\,\mu$s at decode scale,
\S\ref{sec:holder-compute}) add enough fixed overhead that fetch wins
end-to-end below $\Mq$ of a few hundred, \emph{despite} \route's
wire-byte advantage. The wire-byte asymmetry of
\S\ref{sec:select} is necessary but not sufficient at the current host
overhead; three implementation reductions --- a collapsed-response put,
holder-compute amortisation, and cross-request dispatcher batching ---
are what convert it into an end-to-end win, and are engineering, not a
change to the model or the decision. This crossover must not be confused
with the splice tax of \S\ref{sec:bg-prim}: that tax burdens the
\emph{semantic} fetch (move-the-cache-and-adapt), which \route{} avoids
regardless of host overhead. The host-overhead crossover is a property of
our prototype's transport; the splice tax, of the operation itself. The
predicate weighs both.

\subsection{Beyond MLA: compression and selection as one knob}
\label{sec:select-general}
The predicate above consumes byte sizes, not an attention variant, so it
applies wherever cross-instance attention reduces to a set of small
chunks. Three independent levers produce that regime.
\emph{Head-axis compression} (\mla's low-rank latent, the case we
measure) shrinks every token's cache entry. \emph{Token-axis
compression} (the learned $m$-token-to-one compressor in
DeepSeek-V4's CSA~\cite{deepseekv4}) instead shrinks the \emph{number}
of entries; V4 compresses the cache further still, interleaving 4:1 and
128:1 reduction across its layers, which makes a fetched chunk smaller \emph{and} a routed query
relatively heavier, shifting the crossover (\S\ref{sec:select}) without
changing the decision. \emph{Selection} (the top-$k$ Lightning
Indexer of DeepSeek Sparse Attention~\cite{dsa} and the block selection
of NSA~\cite{nsa}) attends to only a few \emph{disjoint} chunks per query, shrinking the
attended set on the entry \emph{count} alone --- independent of how wide
each entry is --- so each query already faces the \route/\fetch/\local{}
choice over a small, scattered set.

\begin{figure*}[t]
\centering
\begin{minipage}[b]{0.49\linewidth}
\centering
\includegraphics[width=\linewidth]{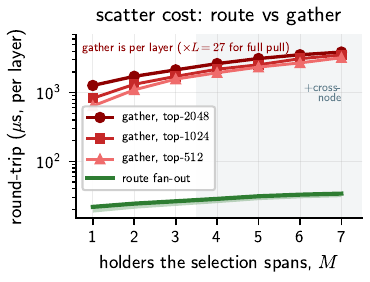}\\
{\footnotesize (a) scatter transport: gather grows with $M$, route fan-out flat}
\end{minipage}\hfill
\begin{minipage}[b]{0.49\linewidth}
\centering
\includegraphics[width=\linewidth]{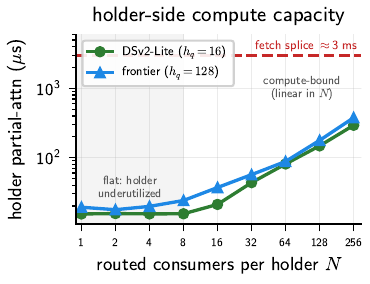}\\
{\footnotesize (b) holder compute: flat to $N{\approx}8$ routed requesters}
\end{minipage}
\Description{(a) Log-scale line plot of round-trip latency versus the number of
holders $M$ a selected set is scattered across: three red curves (top-512/1024/2048
selected entries) rise monotonically with $M$ from a few hundred microseconds to a
few milliseconds per layer, with no kink where holders begin to span a second node;
a green band near the bottom marks the route fan-out, flat at a few tens of
microseconds across all $M$. (b) Log--log plot of holder partial-attention latency
versus the number of routed requesters $N$ for two head counts (16 and 128): both
curves are flat to about $N{=}8$, then rise linearly as compute saturates, both far
below a dashed line at the three-millisecond fetch splice.}
\caption{\textbf{Under selection, route's cost stays flat where the alternatives
grow} (cross-node H100, IBGDA). \textbf{(a)}~Scatter transport. Gathering a $K$-entry
selected set spread across $M$ holders (\fetch{}, red, per layer) grows with $M$ ---
scattering defeats bulk coalescing, so each holder is a separate transfer --- and is
fabric-invariant (no kink as holders cross the node boundary at $M{\ge}4$, shaded);
the full pull scales this by $L{=}27$ layers. The route fan-out (green: single-hop
route-RT plus the $M$-way merge) stays flat at tens of microseconds, shipping the
query once. \textbf{(b)}~Holder compute, the route twin of the $K$-stream copy elbow
(\S\ref{sec:regime}). Measured with FlashMLA's absorbed-\mla{} decode kernel on one H100
($c_t{=}2048$, bf16): a holder serving $N$ routed requesters runs a batched partial
of size $N$, flat to $N{\approx}8$ (GPU underutilised, requesters nearly free) then
linear; at decode ($N{\le}16$) it is $15$--$37\,\mu$s and even saturated ($N{=}256$)
stays $\le0.4$\,ms, far below \fetch's $\approx$3\,ms splice (dashed).}
\label{fig:select-regime}
\end{figure*}

Concretely, these selection budgets are small and now fixed in the released
model configs: DeepSeek-V3.2's DSA indexer selects the top-$2048$ tokens per
query (its \texttt{index\_topk})~\cite{dsa}, GLM-5.1's DSA indexer the same
$2048$~\cite{glm51}, DeepSeek-V4 the top-$1024$ (V4-Pro) or top-$512$ (V4-Flash)
of its compressed entries~\cite{deepseekv4}, and NSA $\approx$16 blocks of 64
($\approx$1024 tokens) plus a 512-token window~\cite{nsa}. All sit at
$512$--$2048$ attended entries, so our $c_t{=}2048$ operating point is exactly
the V3.2/GLM-5.1 selection budget rather than an arbitrary chunk size, and the
$(\Mq,c_t)$ crossover of \S\ref{sec:select} reads as a per-\emph{selected-set}
decision: that same break-even, evaluated at these counts, spans $\approx$270
query rows (V4-Flash's top-$512$) to $\approx$1080 (top-$2048$). Even the
tightest, V4-Flash, stays above a decode batch ($\Mq{\le}256$), so \route{}
wins at decode across the whole family.

Selection
sharpens this beyond the byte count in two ways: the selected set is
\emph{scattered} (disjoint entries, so \fetch{} becomes a gather over
non-contiguous, possibly multi-holder cache, while a routed query stays one
small message), and it is \emph{re-chosen every decode step} (so a fetched set
cannot be amortised the way the static prefix of \S\ref{subsec:rules} can).
We measure the scatter penalty directly on the 2-node testbed
(Fig.~\ref{fig:select-regime}(a)): gathering a
$2048$-entry selected set spread across $M$ holders grows from $\approx\!1.3$ to
$\approx\!3.9$\,ms per layer as $M$ goes $1{\to}7$ (scattering defeats bulk
coalescing, so each holder becomes a separate transfer), whereas the route
fan-out stays flat, the $M$ query sends being probe-bound and the $M$-way
online-softmax merge costing $\le\!25\,\mu$s. \fetch{} is itself fabric-invariant
here, as \route{} is (\S\ref{sec:sens-scale}: same-node and cross-node gathers
track within a few percent), so what \fetch{} pays is the moved bytes and the
per-holder handshakes, not the link.

The asymmetry behind this is structural and
\emph{layer-independent} for \route{}: the requester ships its query once and the
holder runs the $L$-layer partial in place (\S\ref{sec:holder-compute}), while
\fetch{} must pull the selected \ckv{} across all layers ($\approx$64\,MB at
top-$2048$, $L{=}27$, versus a $\sim$2\,KB query). The absolute per-byte gather
rate here is host-copy-bound (\S\ref{sec:overhead}), which inflates the raw
ratio; but the query-versus-cache asymmetry and \route{}'s layer-independence are
what keep \route{} cheapest at decode, and they hold at full wire bandwidth.

The benign-scatter result is not specific to our IB fabric: the knee-free,
linear-in-$M$ gather reproduces on NVSwitch (A100) and on cross-socket PCIe
(A40, RTX~Pro~6000) alike. The $M$-sweep gather is in fact \emph{fabric-insensitive
in shape}: it is serial (the requester pulls holders one at a time) and, in
our prototype, host-copy-bound, so no fabric shows a super-linear knee as holders
cross the socket boundary. The fabric surfaces instead in a \emph{controlled
per-holder} probe: on cross-socket PCIe~Gen4 a fixed-size per-holder gather runs
$\approx$36\% slower across the socket boundary than within it, the penalty
growing with payload, a genuine UPI/CPU-root wire effect that the
non-blocking fabrics do not pay, and that concurrent \emph{route} flows on the
same path also incur (\S\ref{sec:sens-scale}). Either way the per-byte and
per-flow cross-socket penalties fall on \fetch's multi-holder gather, not on
\route's single query, so \route's advantage over \fetch{} only \emph{widens}
where the fabric is weakest.

Seen this way, \route{} \emph{is} the selection step made distributed: the
indexer picks which entries a query attends, and \route{} attends them where
they already reside, shipping the query and merging the partials
(\S\ref{sec:setup-correct}). Native sparse attention attends a disjoint selected
set at the entries' own positions~\cite{nsa,dsa}, so those entries need no
re-rotation: the position-adaptation splice (\S\ref{sec:bg-prim}) is a property
of \emph{contiguous} reuse (a chunk re-inserted at a new offset), not of
selection, and here \route{} wins on the byte asymmetry, scatter, and
re-selection alone.

DeepSeek's own trajectory tracks this: it moved from \mla (head-axis compression)
in V2/V3, to selection layered on \mla in V3.2, to token-axis
compression plus selection in V4, progressively loosening the tie
between the ``small routable chunk'' and \mla's specific latent. We
therefore frame the contribution around the chunk byte budget and
measure the \mla corner, the most widely deployed today; the model's
inputs change for each lever, its structure does not.

\subsection{Serving rules of thumb}
\label{subsec:rules}
The predicate reduces to a few decisions a scheduler can make per
(chunk, request) from quantities it already tracks: the routed-query batch
$\Mq$, the chunk size $c_t$, and the fabric's $(T_{\mathrm{probe}}, BW)$.
\begin{itemize}
\item \textbf{Default to \route{} at decode.} For per-step decode
  redistribution ($\Mq\!\lesssim\!10^3$), \route{} costs tens of microseconds
  ($\approx$31--48\,$\mu$s at $\Mq{=}256$ across the fabrics we measured,
  $60$--$100\times$ below \fetch's $\approx$3\,ms splice). (That is the move-the-cache
  \fetch{}, which always pays the splice; only a \emph{splice-free} bytes-back
  transfer competes with \route{} at decode, and only until the host-overhead
  reductions of \S\ref{sec:overhead} land.) The \route/\fetch{} ranking
  inverts only near $\Mq\!\gtrsim\!10^5$, far above any decode batch; when in
  doubt at decode, route.
\item \textbf{\fetch{} only to amortise.} Moving the cache pays a flat
  $\approx$3\,ms position-adaptation splice up front, so it wins only when the
  pulled \ckv{} will be attended by \emph{many} subsequent local steps on the
  same instance, not for a one-shot cross-instance attention.
\item \textbf{\local{} (re-prefill) only for small chunks.} Re-prefill scales
  with chunk size ($c_t\!\cdot\!L\!\cdot\!c$), so it undercuts the flat splice
  only below $\approx$75--220 tokens (Figure~\ref{fig:crossover}); above that,
  prefer \route{}, falling back to \fetch{} only when no route to the holder
  exists (the disaggregated-prefill regime).
\item \textbf{Contention does not reprice the decision.} \route{} latency is
  flat until a link is fully subscribed; even then ($K{=}3$) it stays
  more than an order of magnitude below the splice (\S\ref{sec:sens-scale}), so a scheduler need only cap concurrent
  flows per link rather than re-rank primitives under load.
\item \textbf{Choose the fabric by probe, not peak bandwidth.} At decode the
  routed payload is too small to exercise peak $BW$, so the operative cost is
  the $\sim$1--5\,$\mu$s probe; an idle PCIe link serves a routed query about as
  well as a busy NVLink one.
\item \textbf{Under sparse selection, route the indexer's choice.} When an
  indexer picks a few \emph{scattered} KV blocks (DeepSeek-V3.2/V4, GLM-5.1),
  \route{} \emph{is} that selection made distributed: the holder attends them in
  place (sparse kernel, tens of $\mu$s, set by the selection budget not the
  partitioned store's size), while moving the cache must \emph{gather} the
  scattered set: a cost that grows with the holders it spans
  ($\approx$3$\times$ from one to seven) and compounds across a socket boundary,
  exactly where \route{} stays flat.
\end{itemize}
These follow directly from the measured constants and hold for the regime we
characterise ($\Mq$ from single digits to thousands, kilo-token chunks,
$25$--$300$\,GB/s fabrics); outside it, re-evaluate the predicate directly.

\section{Characterizing the Device-Initiated RDMA Regime}\label{sec:regime}

\subsection{IBGDA versus CPU-proxy at attention's payload}
We toggle NVSHMEM's transport between device-initiated IBGDA and a
CPU-proxy path on the chunk-prefetch pipeline (the $K{=}8$ staging path
of \S\ref{sec:regime}), holding the workload fixed. The latencies here
are \emph{end-to-end} per-fetch p50 --- dominated by the host prefetch
pipeline (worker dispatch, bf16 staging, landing), hence millisecond-scale
and distinct from the microsecond wire round trip of \S\ref{sec:model};
toggling only the transport isolates its contribution. IBGDA wins
decisively: per-fetch p50 $6.0$\,ms vs $8.4$\,ms ($+40\%$ for proxy) and a
steady-state floor $+53\%$ higher, while wall-clock throughput ties within
run-to-run variance. This is the regime recent transport work did
\emph{not} cover: the proxy-beats-IBGDA result holds for the sub-128\,B
messages of MoE dispatch~\cite{ncclgin,transferengine}, but at attention's
kilobyte-scale per-request payload ($q{+}p{\approx}2$\,KB) the GPU thread
fills the work queue fast enough that removing the host hand-off
dominates. The guideline is a payload-size threshold: above $\sim$1\,KB,
device-initiated RDMA is the right substrate for per-request attention
traffic; below it, a proxy. The throughput tie also localises the binding
constraint at this workload to the host pipeline, not the wire
(\S\ref{sec:overhead}).

\subsection{Holder-side staging: the \texorpdfstring{$K$}{K}-stream pool elbow}
\begin{figure*}[t]
\centering
\begin{minipage}[b]{0.61\linewidth}
\centering
\includegraphics[width=\linewidth]{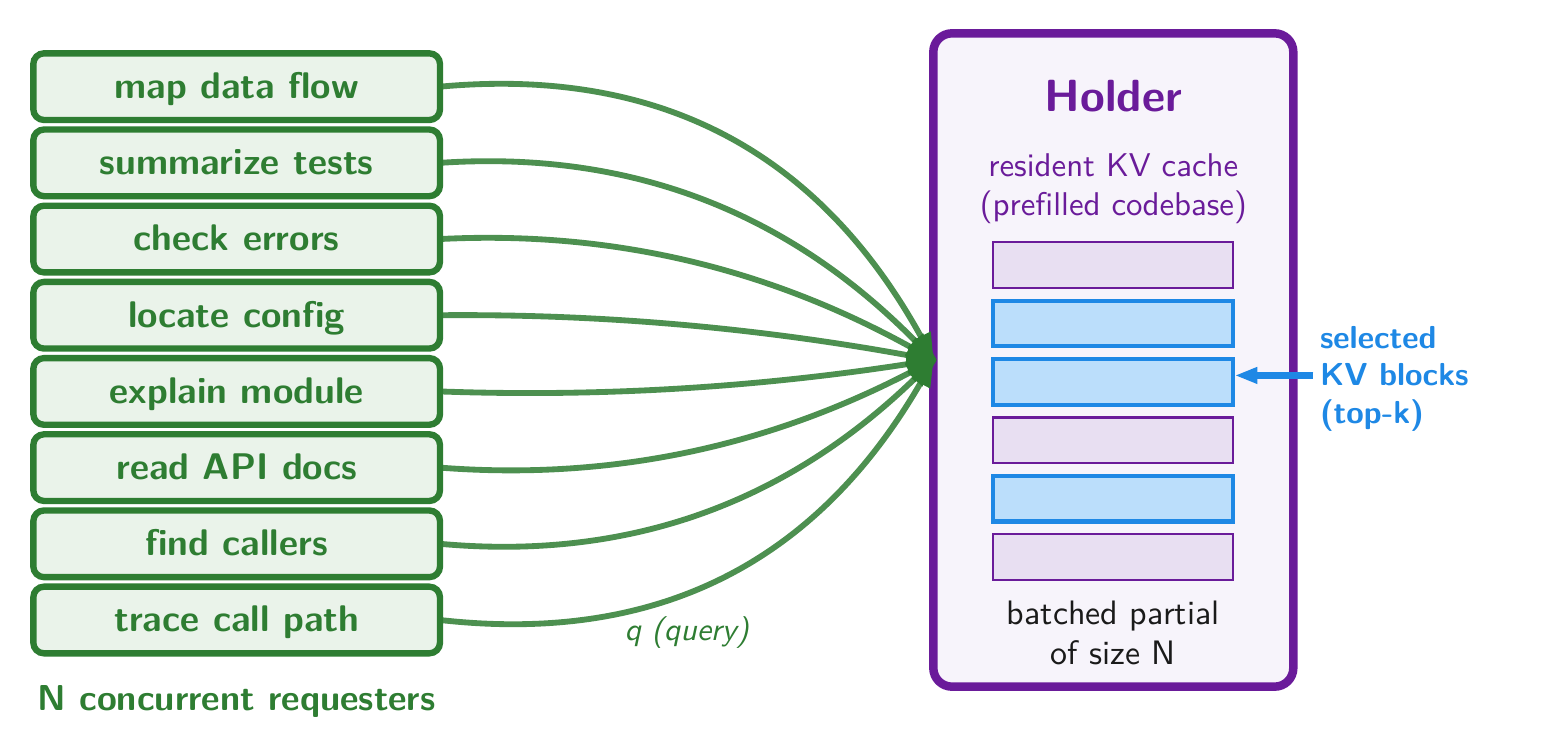}\\
{\footnotesize (a) $N$ routed requesters fan in to one corpus holder}
\end{minipage}\hfill
\begin{minipage}[b]{0.37\linewidth}
\centering
\includegraphics[width=\linewidth]{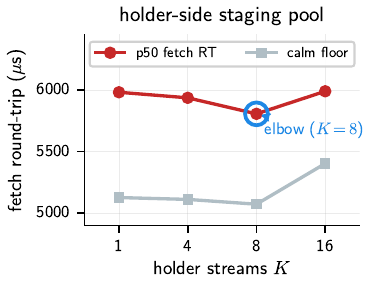}\\
{\footnotesize (b) measured holder-side staging elbow at $K{=}8$}
\end{minipage}
\Description{Left (a): N concurrent routed requesters each send a query to one
holder that owns the resident c-KV corpus and returns a batched partial; a note
marks the copy/compute elbow near N of 8 and a highlighted slab marks the
selected top-k KV block. Right (b): line plot of fetch round-trip latency and a
steady-state floor versus the number of holder-side staging streams K (1, 4, 8,
16), with a knee at K equals 8.}
\caption{\textbf{(a)}~$N$ routed requesters fan in to one corpus holder, which
batches their partials over its resident \ckv{} (the highlighted slab is the
selected top-$k$ block); the copy and compute elbows both sit near
$N{\approx}8$. \textbf{(b)}~Holder-side staging: p50 fetch round trip and
steady-state floor vs.\ the number of CUDA streams $K$ in the holder pool
(chunk-prefetch workload, 2-node $\times$ 4 H100). $K{=}8$ is the elbow;
$K{=}1$ (a single async stream) does not help and $K{=}16$ oversubscribes.}
\label{fig:kstream}
\end{figure*}

The holder stages each incoming request's \ckv into the symmetric heap
via device-to-device copies before the NIC reads them; serialised, these
copies bound throughput, so this staging is where a single holder's capacity
to serve concurrent routed requesters is set. A pool of $K$ CUDA streams
pipelines the copies. We sweep $K\in\{1,4,8,16\}$ (Figure~\ref{fig:kstream}):
$K{=}8$ is the elbow (tail p50 drops $7\%$ and the steady-state floor $9\%$
against the serialised baseline), while $K{=}1$ (async issue on a single
stream) does \emph{not} help, showing the win is HBM-parallel copy engines
rather than mere asynchrony, and $K{=}16$ regresses from scheduler
oversubscription. The serving implication: the elbow is the per-holder fan-out
to target. For a hot canonical chunk (a popular case-law passage or a trending
repository served to many tenants at once), it is the $\approx$8 requester
instances one holder backs before copy-engine contention (rather than the wire)
caps its throughput. This holder-side stream-pool elbow for inter-node RDMA
staging on H100 SXM5 is a concrete deployment knob (set the holder pool to $8$
streams) and one of the implementation reductions \S\ref{sec:select} calls for.

\subsection{Holder-side partial-attention compute: the route-holder's capacity}
\label{sec:holder-compute}
Routing moves the attention \emph{compute} to the holder, so a holder serving
$N$ concurrent routed requesters runs a batched absorbed-\mla{} partial of size
$N$ against its resident \ckv{} and returns each requester's $(o,m,\ell)$ for the
merge. We measure this directly with DeepSeek's production absorbed-\mla{} decode
kernel (FlashMLA~\cite{flashmla}) --- the holder runs no decode pass, but its
partial (a small batch of $N$ query rows over many resident keys) has exactly
that kernel's \emph{shape}, so it is the faithful instrument --- on one H100,
at both the measured DeepSeek-V2-Lite geometry ($h_q{=}16$) and frontier-\mla{}
scale ($h_q{=}128$),
$c_t{=}2048$ (Figure~\ref{fig:select-regime}(b)). The holder's partial-attention
latency is \emph{flat} up to $N{\approx}8$ requesters (the kernel underutilises
the GPU, so extra requesters are nearly free), then turns linear once the GPU
saturates ($N\gtrsim16$). This is the \emph{compute}-capacity twin of the
fetch-holder's \emph{copy}-capacity elbow (\S\ref{sec:regime}): the same
``cap concurrent flows per holder'' rule, set now by compute rather than copy
engines, at a similar fan-out. Holder compute never
approaches the cost it replaces: at the decode operating point ($N\le16$) the
partial is $15$--$37\,\mu$s (comparable to routing's own transport and
$\sim$$100\times$ below \fetch's $\approx$3\,ms splice), and even fully
saturated at $N{=}256$ it is $\le0.4$\,ms, still $\sim$$8\times$ below the
splice. Placing attention compute on the holder, the choice routing makes, is
therefore cheap and bounded; it is what lets a model-serving holder absorb
fan-in without the move-the-cache tax, and it delimits where \route{} applies
(a model-agnostic byte store cannot run the partial and must \fetch{},
\S\ref{sec:select}).

A selection-regime holder runs the \emph{sparse} decode kernel (the indexer's
top-$k$ path) rather than the dense one, and two measured properties carry this
capacity argument into that regime. First, its cost tracks the \emph{selection
budget}, not the store behind it: attending a top-$k$ set costs essentially the
same whether the canonical \ckv{} it draws from is $2$K or $32$K tokens (within
$\sim$15\% across a $16\times$ corpus-size range, $h_q{=}128$), because the
indexed gather touches $k$ entries wherever they reside. A holder's per-query
compute therefore stays flat as the partitioned canonical store \emph{scales}:
the store can grow without inflating the routed query's holder cost, exactly
the property a large partitioned corpus needs. Second, the indexed gather is a
small, bounded premium over the dense-decode kernel at matched $k$: $1.1\times$ at
$k{=}512$, widening to $2$--$3\times$ at $k{=}2048$ as the gather lengthens,
but still $17$--$60\,\mu$s, below \fetch's splice by $\sim$50$\times$ even at the
largest budget, so \route{} stays the cheaper primitive under selection by the
same margin. (Measured on FlashMLA's bf16 sparse kernel; the production FP8
sparse-decode kernel would only narrow the premium.)

This elbow has a direct reading for the agentic workload of \S\ref{sec:intro}:
$N{\approx}8$ is the number of concurrent sub-agents one immutable-prefix holder
serves almost for free before added agents cost linearly and a second replica
(a \fetch{}) is warranted. That replication boundary, not the splice, governs
the pure-prefix case. When a chunk is served at the position it was cached (a
true prefix, shared by every agent at offset~0), \fetch's $\delta$-rotation is
the identity and elides, so \fetch{} is at its \emph{cheapest} and the
move-the-cache splice no longer separates the primitives. \route{} still wins
per agent there, but on the decode byte-asymmetry (a kilobyte-scale query versus
the whole document's \ckv) and on sparing a full document pull each time a
sub-agent lands on a fresh instance, not on the splice; \fetch{} overtakes only
once enough agents co-locate on one replica to amortise that pull
(\S\ref{subsec:rules}). The prefix case is thus where the \emph{full} predicate,
not the splice alone, earns its keep.

\section{Sensitivity: Payload Geometry}\label{sec:sens-geom}
Two payload-geometry sensitivities bound the predicate's inputs, and they
have opposite shapes. \emph{Route} is linear in the query batch:
$T_{\mathrm{route}}$ holds near its $\approx$16\,$\mu$s probe floor for
$\Mq\le128$ (fixed cost dominates), then rises through the per-byte
regime ($\approx$116\,$\mu$s at $\Mq{=}1024$, $\approx$388\,$\mu$s at $\Mq{=}4096$ for
the real \mla payload) at the payload-independent $\approx$25\,GB/s
slope of Table~\ref{tab:payload}. \emph{Splice}, the \fetch-side cost,
has the complementary geometry: it is \emph{flat} in chunk size, measured
at $2.77/2.78/2.91/3.06$\,ms for $c_t{=}55/1024/2048/4096$: only
$\sim$10\% growth over a $74\times$ token range, because the per-layer
$\delta$-rotation that dominates it ($\approx$80\%) is launch-bound, not
token-bound (the gather stays flat at $\approx$320\,$\mu$s; only the
scatter begins to scale past $c_t{=}1024$). The consequence is a clean
division of labour for the predicate: \route's cost is set by \emph{how
many queries} attend a chunk, \fetch's by \emph{almost nothing} (a fixed
$\approx$3\,ms), and \local's by \emph{how many tokens} the chunk holds.

\section{Sensitivity: Scaling and Topology}\label{sec:sens-scale}
\paragraph{Fabric and scope.} The probe term $T_{\mathrm{probe}}(F,s)$ is
fabric- and scope-specific. Our device-side sweep measures intra-NVL
(NVLink) block-scope at $\approx$1\,$\mu$s and cross-NVL IBGDA block-scope at
$\approx$16\,$\mu$s (the put-plus-signal primitive of \S\ref{sec:setup}; the
older GET-based round trip was $26\,\mu$s), with warp- and thread-scope
progressively slower. The model carries these as distinct
$T_{\mathrm{probe}}$ values, which is why it places the route/fetch/local
boundary differently per fabric: inside an NVLink island a
$\sim$1\,$\mu$s probe (measured $1.2$--$1.6\,\mu$s for P2P over NVLink, a higher
$\approx$5--9\,$\mu$s over intra-node PCIe) makes even fine-grained
redistribution viable, whereas across islands the $\approx$16\,$\mu$s IBGDA
probe sets the $\Mq$ at which routing's per-byte term starts to matter. This
cross-NVL cost is, moreover, \emph{flat across the fat-tree}: a spine-traversing
cross-leaf node pair measures nearly the same probe ($15.9$ vs $14.6\,\mu$s) and
round trip ($119.9$ vs $114.6\,\mu$s at $\Mq{=}1024$, $\approx$25\,GB/s either
way) as a same-leaf pair, so routing's cost turns on the intra- versus cross-NVL
boundary, not on which leaf holds the \ckv.

\begin{figure}[t]
\centering
\includegraphics[width=\linewidth]{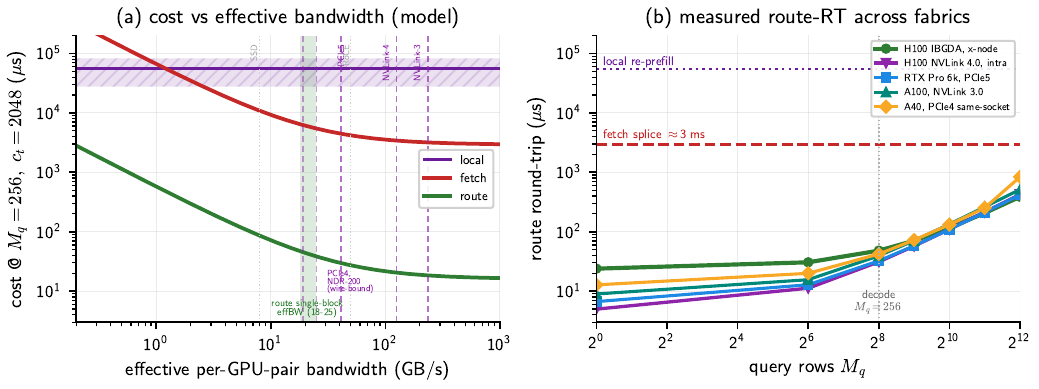}
\Description{Two-panel figure. Panel (a): a log--log model plot of
redistribution cost versus fabric bandwidth from 0.2 to 1000 GB/s at a fixed
decode operating point; route declines with bandwidth and stays well below
fetch's flat 3 ms splice line and local's flat re-prefill band across the
marked SSD, RoCE, PCIe, and NVLink~4.0 bandwidths. Panel (b): measured route
round-trip versus query rows for five fabrics (A40 PCIe~Gen4, A100 NVLink~3.0,
intra-node H100 NVLink~4.0, RTX~Pro~6000 PCIe~Gen5, and cross-node H100 IBGDA); the five curves nearly
coincide at the decode batch around thirty to fifty microseconds, far below
the fetch splice and local re-prefill reference lines, and fan out only at the
largest batches.}
\caption{Fabric robustness of redistribution at the decode operating point
($\Mq{=}256$, $c_t{=}2048$, DeepSeek-V2-Lite). \textbf{(a)~Model:}
\route{} (green) scales as $\sim$1/$BW$ and stays cheapest across four orders
of magnitude; \fetch{} (red) floors at its $\approx$3\,ms splice above SSD-tier
$BW$ (its all-layer pull dominates only below that) and \local{} (purple) is
re-prefill-compute-bound. The shaded green band is the \emph{measured} single-block route effBW across all five fabrics ($18$--$25$\,GB/s, what \route{} sees); the dashed ticks are the \emph{measured per-GPU-pair} link rates (what \fetch's bulk pull sees) --- PCIe~Gen4 $19$\,GB/s and NDR-200 $25$\,GB/s fall in-band (wire-bound), while PCIe~Gen5 $41$\,GB/s, the H100 NVLink~4.0 (NV6 direct) $\approx$125\,GB/s, and the A100 NVLink~3.0 (NVSwitch) $235$\,GB/s sit to its right, the dispatch headroom routing never uses. \textbf{(b)~Measurement:} \route{} round trip
versus $\Mq$ on five real fabrics: A40 PCIe~Gen4 (same-socket), A100 NVLink~3.0,
intra-node H100 NVLink~4.0, RTX~Pro~6000 PCIe~Gen5, and cross-node H100 IBGDA. At decode ($\Mq{=}256$) the
five cluster at $\approx$31--48\,$\mu$s and stay over $60\times$ below \fetch's
splice: the single-block dispatch is capped at $\approx$20\,GB/s, so route-RT tracks
single-block \emph{dispatch} throughput, not fabric peak --- the same H100's
$900$\,GB/s NVLink~4.0 sustains only $\approx$21\,GB/s for one block, a hair below
its own cross-node IBGDA ($\approx$25$\,$GB/s) --- so at decode it is effectively
fabric-invariant across the five single-bottleneck fabrics.
Bandwidth separates the fabrics only in the large-batch tail.}
\label{fig:bwsens}
\end{figure}

\paragraph{Fabric bandwidth.} The probe varies with fabric, and so does the
per-byte term through $BW(F)$; but at fabric bandwidth \fetch's cost is
splice-dominated (its all-layer pull is a smaller, $BW$-scaling term) and
\local's is compute-bound, so \route's transport scales with $BW$ the most and
stays cheapest; the \emph{winner}, not merely the latency, is
fabric-robust. Figure~\ref{fig:bwsens}(a) sweeps the model from $0.2$ to
$10^3$\,GB/s at the decode point ($\Mq{=}256$, $c_t{=}2048$): \route{} stays
one-to-three orders of magnitude below \fetch{} and \local{} across SSD, RoCE,
PCIe, and NVLink alike, \fetch{} flooring at its $\approx$3\,ms splice above
SSD-tier bandwidth. Figure~\ref{fig:bwsens}(b) confirms this on hardware:
the \emph{measured} \route{} round trip on five fabrics spanning PCIe~Gen4,
NVLink~3.0 and~4.0, PCIe~Gen5, and cross-node IBGDA --- a
$>$$10\times$ span of nominal bandwidth across two transports (intra-node P2P and
cross-node RDMA). At the decode batch the five single-bottleneck fabrics cluster
within $1.5\times$ ($\approx$31--48\,$\mu$s at $\Mq{=}256$), all over $60\times$
below \fetch's $\approx$3\,ms splice.

A decode-sized routed dispatch carries so little that a \emph{single} thread
block issues the whole transfer, and one block can only push puts into the
link at $\approx$18--25\,GB/s --- well under most of these fabrics' peaks. That
issue rate, not the wire, is the bottleneck, so the dispatch is
\emph{dispatch-bound, not bandwidth-bound}: the transfer term is nearly the
same on every fabric, and the only fabric-specific cost left is the small
probe. Peak bandwidth then does not even \emph{order} the fabrics at decode.
The A100 on NVLink~3.0 ($39\,\mu$s at $\Mq{=}256$) is \emph{slower} than
the RTX~Pro~6000 on the nominally $\sim$6$\times$ lower-peak
PCIe~Gen5 ($32\,\mu$s): at one block neither approaches its ceiling, so the
faster wire wins nothing. Most pointedly, the \emph{same} H100 routes over its
own NVLink~4.0 (a $900$\,GB/s per-GPU mesh, aggregated over all $18$ links)
at only $\approx$21\,GB/s for a single block, a hair \emph{below} its
cross-node IBGDA ($\approx$25\,GB/s): to a one-block dispatch a local NVLink
mesh and a cross-node RDMA NIC look alike. NVLink~4.0 leads only out to
$\Mq{\approx}1000$, on its $\approx$1\,$\mu$s probe, before the InfiniBand path
(whose RDMA engine issues a touch faster) takes over.

A dedicated multi-block put-bandwidth benchmark settles that the wire is not
the limit: driven by many blocks it recovers each link's true peak, and those
peaks span more than $10\times$. Yet a single-block route saturates the wire on
only the \emph{slowest} fabric, and leaves a wider and wider margin unused as
the wire speeds up: on the A40's PCIe~Gen4 its $18.7$\,GB/s already $\approx$
the $19$\,GB/s peak (wire-bound); on the RTX's PCIe~Gen5 it draws about half
($22$ of $41$\,GB/s); on the H100's own NVLink~4.0 under a fifth ($21$ of $\approx$125\,GB/s,
six direct links and no switch); on the A100's even-wider NVLink~3.0 under a
tenth ($18$ of $235$\,GB/s, fanned through an NVSwitch). The ladder \emph{is} the
intuition: a single block is wire-bound only where the wire itself is
$\approx$20\,GB/s, and dispatch-bound on everything faster. The measured
$\approx$18--25\,GB/s rates are dispatch ceilings, not link ceilings, so routing
never reaches for NVLink's headroom; peak re-asserts only in the large-batch
tail (the A40 same-socket super-linear regime past $\Mq{=}2048$).

An isolated A40 \emph{cross-socket} flow can read far slower
($0.7$--$1.1$\,ms at $\Mq{=}1024$), but that number is non-reproducible and
\emph{vanishes under load}: with any concurrent cross-socket traffic it settles
to $\approx$131\,$\mu$s (its same-socket cost), so we attribute it to an
idle-link warm-up effect on an otherwise-quiescent path (mechanism not isolated)
and report the warm value, not a steady worst case.

A block-count sweep shows this single-block rate is \emph{issue}-bound, not a
hardware ceiling: on intra-node NVLink it scales $\approx$6$\times$ ($\approx$21
to $\approx$125\,GB/s) with more blocks, matching NVIDIA's NVSHMEM
put-bandwidth benchmark and the $6$-link (\texttt{NV6}) GPU-pair ceiling. So
multi-block dispatch \emph{would} shrink the transfer term (including at
decode, where it dominates), but two things make that headroom moot for the
deployed route. First, redistribution is \emph{point-to-point}: it rides one GPU
pair's link and never the $\approx$900\,GB/s per-GPU aggregate (which fans across
all $18$ NVLinks to all peers), so the real headroom is the $\approx$6$\times$,
not $\approx$40$\times$. Second, at decode the single-block route already sits
$>$$60\times$ below \fetch, so the extra bandwidth buys latency the
route-vs-\fetch{} decision does not need, and each added block costs an SM the
co-located holder spends on its attention compute (\S\ref{sec:holder-compute}).
Single-block dispatch is thus the natural minimal-footprint operating point, and
multi-block parallelism would matter only in the large-$\Mq$ transfer tail,
the regime where \fetch{} already wins. The \route/\fetch{} crossover is set
by \fetch's compute splice and the host overhead of \S\ref{sec:overhead}, not by
\route's byte term, so the block count does not move it.

\fetch{} is the mirror image: its bulk \ckv{} pull
\emph{is} bandwidth-bound (we model it as a coalesced bulk transfer, not a
single block), yet \fetch{} is splice-\emph{compute}-dominated, so even an
instantaneous pull would leave it at its $\approx$3\,ms splice floor. The
route-vs-\fetch{} verdict is therefore invariant to how many blocks either
primitive issues: latency binds the one we deploy, compute the other.


\paragraph{Congestion.} Our latencies are measured on an otherwise-idle
fabric; a production network is contended. Two of congestion's effects are
already on axes we characterise. (i)~It lowers \emph{effective} bandwidth:
Figure~\ref{fig:bwsens}(a) shows \route{} stays cheapest as $BW$ falls to
$0.2$\,GB/s; equivalently, \route{} loses to \fetch{} only once effective
bandwidth drops below $\Mq(q{+}p)/T_{\mathrm{splice}}\approx0.2$\,GB/s at
$\Mq{=}256$ (a $\sim$125$\times$ degradation from $25$\,GB/s), because
\fetch's $\approx$3\,ms splice and \local's re-prefill are congestion-immune
\emph{compute}. (ii)~It adds queueing to the probe, an additive term: even a
$10\times$ probe inflation leaves \route{} an order of magnitude below
\fetch. The route-vs-\fetch{} \emph{ranking} is thus robust to congestion by
cost structure; only \route's absolute latency rises.

\begin{figure}[t]
\centering
\includegraphics[width=\linewidth]{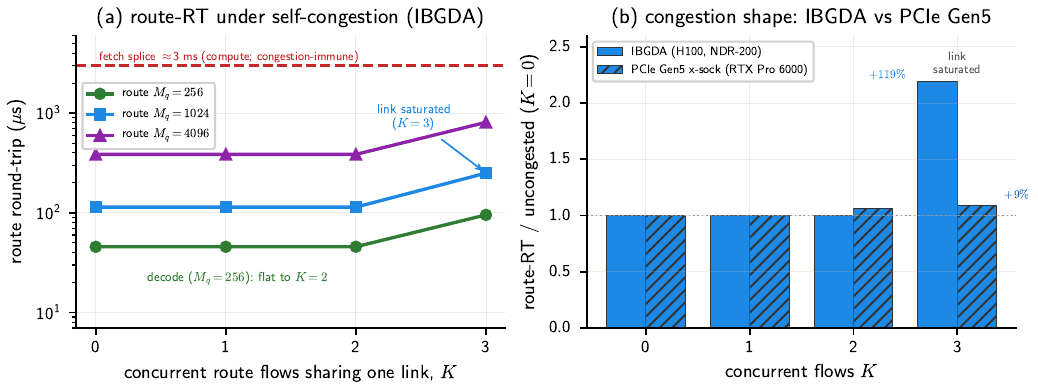}
\Description{Two-panel figure. Panel (a): a line plot of route round-trip versus
the number of concurrent flows $K$ sharing one link, for three query-batch
sizes; the decode-sized batch ($M_q{=}256$) stays flat across load, only the
largest batch rises at $K{=}3$ (link saturated), and all curves remain far below
a dashed reference line marking fetch's 3 ms splice. Panel (b): a grouped bar
chart of route round-trip at $M_q{=}1024$ normalized to the uncongested ($K{=}0$)
baseline, comparing IBGDA (H100, solid bars) with cross-socket PCIe Gen5 (RTX Pro
6000, hatched bars); both stay at 1.0 through $K{=}2$ and rise only at
$K{=}3$, IBGDA by 119 percent and PCIe Gen5 by 9 percent.}
\caption{\route{} round trip under self-congestion. \textbf{(a)}~$K$ concurrent
route flows share one NDR-200 link (two same-leaf H100 nodes); latency is flat
through $K{=}2$ at every batch and rises only once the link is fully subscribed
($K{=}3$; $\Mq{=}1024$: $114{\to}250\,\mu$s, $+119\%$), every case far below
\fetch's $\approx$3\,ms splice (dashed); the route-vs-\fetch{} ranking never
inverts. \textbf{(b)}~The \emph{same} flat-until-saturation shape reproduces on an
unrelated fabric: route-RT at $\Mq{=}1024$ normalized to $K{=}0$, on IBGDA
(H100, NDR-200) versus cross-socket PCIe~Gen5 (RTX~Pro~6000). The
bandwidth-tight NDR-200 link congests hardest ($+119\%$), the higher-headroom
PCIe~Gen5 barely ($+9\%$); a cross-socket A40 PCIe~Gen4 flow shows the same
$K{=}3$ onset ($+36\%$ vs its warm baseline) and an A40 \emph{same}-socket
control stays flat, so the rise tracks shared-link subscription, not the GPU.}
\label{fig:congestion}
\end{figure}

We confirm this
empirically (two same-leaf H100 nodes, $K$ concurrent route flows sharing one
NDR-200 link, the requester's NIC uplink; Figure~\ref{fig:congestion}): the measured route round trip is
\emph{flat} through $K{=}0$--$2$ (probe $14.5\,\mu$s, $\Mq{=}256$ at
$45.6\,\mu$s, unchanged), and rises only once the link is fully subscribed at
$K{=}3$, then across the board, the queueing landing on the probe as much as
the transfer (probe $14.5{\to}39.5\,\mu$s; $\Mq{=}256$ $45.6{\to}95$;
$\Mq{=}1024$ $114{\to}250\,\mu$s). Even fully congested, that $\Mq{=}1024$ round
trip stays $\sim$12$\times$ below \fetch's splice, so the route-vs-\fetch{}
ranking never inverts.

This congestion law is not InfiniBand-specific: it
reproduces on cross-socket PCIe ($K$ concurrent route flows sharing an A40 node's
PCIe/UPI path), where $K\le2$ stays flat and the fully subscribed $K{=}3$ flow
rises $+36\%$ ($131{\to}178\,\mu$s at $\Mq{=}1024$, against the warm baseline),
the same flat-until-saturation shape as IB; a PCIe~Gen5 node (RTX~Pro~6000) shows
the same onset more gently ($+9\%$), its higher per-flow bandwidth softening the
saturation, while an A40 \emph{same}-socket control stays flat across $K$ (the
PCIe switch isolates same-socket P2P), placing the effect on the shared
cross-socket path, not the GPUs (Figure~\ref{fig:congestion}b). And of the two
\emph{network} primitives \route{} loads the fabric least (fewest bytes
per redistribution, \S\ref{sec:bg-transport}), so it is the congestion-friendly
choice (\local{} touches no network at all, but pays the full re-prefill). What remains is many
\emph{concurrent} redistributions contending for one holder or link: the
holder-side $K$-stream staging of \S\ref{sec:regime} bounds the former, and
scheduling the latter across tenants is the serving layer's task, which our
predicate feeds.

\paragraph{Tensor parallelism.} Under TPLA~\cite{tpla} at degree $N$ the
latent is column-partitioned across ranks. Cross-instance routing then
pairs ranks: $A.\text{rank}_r$ ships its $\Mq\times d_{qk}/N$ query slice
to $B.\text{rank}_r$, and the cross-rank all-reduce stays inside each
instance over NVLink. Per-rank inter-node bytes fall by $1/N$ ($50\%$ at
$N{=}2$, measured); the aggregate is unchanged but $N$ rank-pairs proceed
in parallel, so routing scales \emph{with} tensor parallelism rather than
against it. This \emph{cross-instance} rank-pairing (across two independent instances,
unlike Helix's rank-paired merge \emph{within} one deployment~\cite{helix}) has
not, to our knowledge, been characterised.

\section{Related Work}\label{sec:related}
\paragraph{Routing the query.}
Shipping queries to the instance that holds the keys/values, computing
partial attention remotely, and merging via online softmax was
introduced by DistAttention~\cite{distattention} for sharded KV within a
single cluster and extended to \mla within one tensor-parallel deployment
by Helix~\cite{helix}; Meta's context-parallel \emph{pass-Q}~\cite{cp_meta}
applies it within one long-context job, and
Adrenaline~\cite{adrenaline} makes the same query$\rightarrow$KV choice
to offload decode attention onto prefill instances. We adopt the merge
unchanged~\cite{onlinesoftmax,flashattention} and claim no novelty for
routing itself. Our setting differs on three axes none of these address:
a \emph{cross-instance, partitioned canonical store} (not a single
cluster, a single job, or role-specialised prefill/decode), the \emph{\mla
economic regime} that makes routing fine-grained-viable, and a
\emph{device-initiated} RDMA substrate.

\paragraph{Cross-instance KV migration.}
Mooncake~\cite{mooncake}, DistServe~\cite{distserve}, and
LMCache~\cite{lmcache} disaggregate prefill from decode and migrate \ckv
across nodes over RDMA; BanaServe~\cite{banaserve} adds attention-level \ckv{}
migration and a shared global KV store to rebalance disaggregated load. Their
regime is role-specialised (the decode worker has no route to the prefill state
except to receive it), so the cache \emph{must} move.
ServerlessLLM~\cite{serverlessllm} migrates a running request by shipping its
\emph{tokens} and recomputing KV at the destination rather than transferring the
cache --- our \local{}, chosen there because full KV is too large to move; \mla's
compressed latent revives \fetch, and \route{} has no analogue in that design.
ContextPilot~\cite{contextpilot} instead raises the hit rate of KV that
is \emph{already local} by reordering requests onto a shared prefix, sidestepping
relocation entirely. Ours is cross-peer content distribution with canonical
overlap, where we quantify when moving the cache (\fetch) loses to routing the query.

\paragraph{Position-independent caching and the cost of adaptation.}
Reusing a cached chunk at a position other than where it was computed
requires adapting it. PromptCache~\cite{promptcache} precomputes chunks
against a dummy prefix; CacheBlend~\cite{cacheblend} blends multiple
reused chunks by selectively recomputing $5$--$18\%$ of tokens; and
EPIC~\cite{epic} makes caching position-independent on standard (GQA/MHA)
models, recomputing only a small \emph{carved} prefix per chunk to repair
the attention sink at chunk boundaries, MEPIC~\cite{mepic} reducing it to a
chunk's first block. This adaptation is precisely the
\fetch-side cost our predicate weighs: on \mla it is a $\delta$-rotation
that re-aligns a chunk's decoupled-RoPE band to the target offset
(\S\ref{sec:bg-prim}). We propose no new caching scheme; we characterise
when the adaptation cost (EPIC's carve on GQA, the $\delta$-rotation
splice on \mla) makes \route{} the better primitive, tying this work
to the architecture-agnostic view of \S\ref{sec:select-general}.

\paragraph{Device-initiated RDMA.}
DeepEP~\cite{deepep} uses NVSHMEM IBGDA for MoE all-to-all;
TransferEngine~\cite{transferengine} adds a host-proxy fallback where IBGDA is
unavailable; and NCCL~GIN~\cite{ncclgin} and the portable expert-parallel engines
UCCL-EP and NCCL~EP~\cite{uccl_ep,nccl_ep} report IBGDA/proxy/GDAKI small-message
latencies. All target MoE or collective traffic with sub-kilobyte messages; we
characterise the \emph{attention} regime (kilobyte-scale query batches on
NDR-200\,G) and the holder-side staging it requires (\S\ref{sec:regime}), which
that literature does not.

\paragraph{MLA, its variants, and sparse selection.}
MLA~\cite{deepseekv2} and its tensor-parallel form TPLA~\cite{tpla} (on
which our rank-paired routing builds, \S\ref{sec:sens-scale}) define the
head-axis compression case we measure; DeepSeek Sparse
Attention~\cite{dsa}, DeepSeek-V4's token-axis CSA~\cite{deepseekv4}, and
NSA~\cite{nsa} add selection and token-axis compression, and
IndexCache~\cite{indexcache} reuses an indexer's selected blocks \emph{across
layers} to amortise selection within an instance. These set the
byte budgets our predicate consumes (top-$2048$ selected entries for V3.2 and
GLM-5.1, $512$--$1024$ for DeepSeek-V4; \S\ref{sec:select-general}); none
addresses how the selected chunks move \emph{between} instances. A systems
analysis of \mla/MoE inference~\cite{newbottleneck} likewise pins the interconnect
as the post-\mla bottleneck, but for MoE all-to-all, not the inter-instance KV
redistribution we model.

\paragraph{Topology-aware characterisation.}
The roofline model~\cite{roofline} relates compute to bandwidth on a
single device; the closest published characterisation in spirit is that
of multi-chip GPU data sharing~\cite{multichip}, whose findings on
inter-chip bandwidth non-uniformity seeded a sharing-aware cache
design~\cite{sac}. MoE-CAP~\cite{moecap} benchmarks sparse-MoE serving (S-MBU/S-MFU
utilisation) but holds the attention/KV term fixed: the very term we
characterise, treating the larger expert all-to-all as orthogonal. \emph{Cross-instance attention redistribution} has not, to our
knowledge, been modeled; our probe/transfer/compute/return/merge
decomposition (\S\ref{sec:model}) is that model.

\section{Conclusion}\label{sec:concl}
We set out to answer a transport question (for cross-instance \mla
attention, move the query or move the cache?) and to answer it with a
model, not one system's numbers. The answer is a cost-shape argument:
\fetch{} pays either a flat $\approx$3\,ms splice (contiguous reuse) or an
ms-scale scattered gather (sparse selection), and \local{} a size-scaling
re-prefill, while \route{} pays none of these: on real H100 IBGDA a routed
round trip is $\approx$116\,$\mu$s at $\Mq{=}1024$, one-to-two orders of
magnitude below those alternatives. \mla's narrow latent makes the routed
query small enough that at the small batches typical of decode it also moves
the fewer wire bytes ($\ge$76\% fewer at $\Mq{=}256$). We package this
as a topology-aware cost model (probe $\approx$16\,$\mu$s, payload-independent
$\approx$25\,GB/s, fit to $\approx$7\% in the amortised regime) and a closed-form
\route/\fetch/\local{} predicate whose inputs we characterise across
payload, fabric, and tensor-parallel degree.

Its structure is unchanged in
the regime already deployed: where a sparse-attention indexer
(DeepSeek-V3.2, V4, GLM-5.1) shrinks each query's attention to a few
scattered KV blocks, \route{} \emph{is} that selection made
distributed: the query attends the chosen entries where they already
reside, with no cache re-rotation, a merge we verify exact to bf16 noise on
the production sparse-attention kernel itself
(\S\ref{sec:setup-correct},~\S\ref{sec:select-general}). The model and
predicate are the reusable artefact: together they make a partitioned
canonical \ckv store (or a single large immutable document fanned out to many
concurrent agents) \emph{schedulable}.

For a practitioner the guidance collapses
to a handful of rules over quantities a scheduler already tracks: default to
\route{} at decode (tens of microseconds, $\sim$60--100$\times$ below the
move-the-cache splice); \fetch{} only to amortise a chunk over many subsequent
local steps; cap concurrent flows per holder near the $\approx$8 where both its
copy- and compute-elbows sit; and choose the fabric by probe latency, since a
decode-sized query cannot exercise peak bandwidth, counting on \route's margin
being \emph{largest} where the fabric is weakest: a scattered selection makes
\fetch{} gather a full chunk from every holder, paying a per-holder
cross-socket penalty on PCIe that non-blocking fabrics (IB, NVSwitch) avoid,
while \route{} ships only a small query per holder. The remaining gap is end-to-end:
at our prototype's host overhead the routing transport needs the three
implementation reductions of \S\ref{sec:select} before its wire-byte win
becomes a wall-clock win, and closing it inside a live serving stack
is the natural next step, in the same characterise-then-build spirit that
carried multi-chip data-sharing characterisation~\cite{multichip} into a
sharing-aware cache design~\cite{sac}.

\begin{acks}
This work has been funded by the Free State of Bavaria in the DSgenAI project 
(Grant Nr.: RMF-SG20-3410-2-18-4). 
The authors gratefully acknowledge the scientific support and HPC resources 
provided by the Erlangen National High Performance Computing Center (NHR@FAU) 
of the Friedrich-Alexander-Universität Erlangen-Nürnberg (FAU). 
The hardware is funded by the German Research Foundation (DFG).

\end{acks}

\bibliographystyle{ACM-Reference-Format}
\bibliography{refs}


\begin{thebibliography}{43}


\ifx \showCODEN    \undefined \def \showCODEN     #1{\unskip}     \fi
\ifx \showISBNx    \undefined \def \showISBNx     #1{\unskip}     \fi
\ifx \showISBNxiii \undefined \def \showISBNxiii  #1{\unskip}     \fi
\ifx \showISSN     \undefined \def \showISSN      #1{\unskip}     \fi
\ifx \showLCCN     \undefined \def \showLCCN      #1{\unskip}     \fi
\ifx \shownote     \undefined \def \shownote      #1{#1}          \fi
\ifx \showarticletitle \undefined \def \showarticletitle #1{#1}   \fi
\ifx \showURL      \undefined \def \showURL       {\relax}        \fi
\providecommand\bibfield[2]{#2}
\providecommand\bibinfo[2]{#2}
\providecommand\natexlab[1]{#1}
\providecommand\showeprint[2][]{arXiv:#2}

\bibitem[Bai et~al\mbox{.}(2026)]%
        {indexcache}
\bibfield{author}{\bibinfo{person}{Yushi Bai}, \bibinfo{person}{Qian Dong},
  \bibinfo{person}{Ting Jiang}, \bibinfo{person}{Xin Lv},
  \bibinfo{person}{Zhengxiao Du}, \bibinfo{person}{Aohan Zeng},
  \bibinfo{person}{Jie Tang}, {and} \bibinfo{person}{Juanzi Li}.}
  \bibinfo{year}{2026}\natexlab{}.
\newblock \bibinfo{title}{{IndexCache}: Accelerating Sparse Attention via
  Cross-Layer Index Reuse}.
\newblock
\showeprint[arxiv]{2603.12201}~[cs.CL]
\urldef\tempurl%
\url{https://arxiv.org/abs/2603.12201}
\showURL{%
\tempurl}


\bibitem[Bhatia et~al\mbox{.}(2025)]%
        {helix}
\bibfield{author}{\bibinfo{person}{Nidhi Bhatia}, \bibinfo{person}{Ankit More},
  \bibinfo{person}{Ritika Borkar}, \bibinfo{person}{Tiyasa Mitra},
  \bibinfo{person}{Ramon Matas}, \bibinfo{person}{Ritchie Zhao},
  \bibinfo{person}{Maximilian Golub}, \bibinfo{person}{Dheevatsa Mudigere},
  \bibinfo{person}{Brian Pharris}, {and} \bibinfo{person}{Bita~Darvish
  Rouhani}.} \bibinfo{year}{2025}\natexlab{}.
\newblock \bibinfo{title}{Helix Parallelism: Rethinking Sharding Strategies for
  Interactive Multi-Million-Token {LLM} Decoding}.
\newblock
\showeprint[arxiv]{2507.07120}~[cs.DC]
\urldef\tempurl%
\url{https://arxiv.org/abs/2507.07120}
\showURL{%
\tempurl}


\bibitem[Cheng et~al\mbox{.}(2025)]%
        {lmcache}
\bibfield{author}{\bibinfo{person}{Yihua Cheng}, \bibinfo{person}{Yuhan Liu},
  \bibinfo{person}{Jiayi Yao}, \bibinfo{person}{Yuwei An},
  \bibinfo{person}{Xiaokun Chen}, \bibinfo{person}{Shaoting Feng},
  \bibinfo{person}{Yuyang Huang}, \bibinfo{person}{Samuel Shen},
  \bibinfo{person}{Kuntai Du}, {and} \bibinfo{person}{Junchen Jiang}.}
  \bibinfo{year}{2025}\natexlab{}.
\newblock \showarticletitle{{LMCache}: An Efficient {KV} Cache Layer for
  Enterprise-Scale {LLM} Inference}.
\newblock \bibinfo{journal}{\emph{arXiv preprint arXiv:2510.09665}}
  (\bibinfo{year}{2025}).
\newblock


\bibitem[Dao et~al\mbox{.}(2022)]%
        {flashattention}
\bibfield{author}{\bibinfo{person}{Tri Dao}, \bibinfo{person}{Daniel~Y. Fu},
  \bibinfo{person}{Stefano Ermon}, \bibinfo{person}{Atri Rudra}, {and}
  \bibinfo{person}{Christopher R{\'e}}.} \bibinfo{year}{2022}\natexlab{}.
\newblock \showarticletitle{Flash{A}ttention: Fast and Memory-Efficient Exact
  Attention with {IO}-Awareness}. In \bibinfo{booktitle}{\emph{Advances in
  Neural Information Processing Systems (NeurIPS)}}.
\newblock


\bibitem[{DeepSeek-AI}(2026)]%
        {deepseekv4}
\bibfield{author}{\bibinfo{person}{{DeepSeek-AI}}.}
  \bibinfo{year}{2026}\natexlab{}.
\newblock \bibinfo{title}{{DeepSeek-V4}: Towards Highly Efficient Million-Token
  Context Intelligence}.
\newblock
\urldef\tempurl%
\url{https://huggingface.co/deepseek-ai/DeepSeek-V4-Pro/blob/main/DeepSeek_V4.pdf}
\showURL{%
\tempurl}
\newblock
\shownote{Technical report}.


\bibitem[{DeepSeek-AI} et~al\mbox{.}(2024)]%
        {deepseekv2}
\bibfield{author}{\bibinfo{person}{{DeepSeek-AI}} {et~al\mbox{.}}}
  \bibinfo{year}{2024}\natexlab{}.
\newblock \bibinfo{title}{{DeepSeek-V2}: A Strong, Economical, and Efficient
  Mixture-of-Experts Language Model}.
\newblock
\showeprint[arxiv]{2405.04434}~[cs.CL]
\urldef\tempurl%
\url{https://arxiv.org/abs/2405.04434}
\showURL{%
\tempurl}


\bibitem[{DeepSeek-AI} et~al\mbox{.}(2025)]%
        {dsa}
\bibfield{author}{\bibinfo{person}{{DeepSeek-AI}} {et~al\mbox{.}}}
  \bibinfo{year}{2025}\natexlab{}.
\newblock \bibinfo{title}{{DeepSeek-V3.2}: Pushing the Frontier of Open Large
  Language Models}.
\newblock
\showeprint[arxiv]{2512.02556}~[cs.CL]
\urldef\tempurl%
\url{https://arxiv.org/abs/2512.02556}
\showURL{%
\tempurl}


\bibitem[Fu et~al\mbox{.}(2024)]%
        {serverlessllm}
\bibfield{author}{\bibinfo{person}{Yao Fu}, \bibinfo{person}{Leyang Xue},
  \bibinfo{person}{Yeqi Huang}, \bibinfo{person}{Andrei-Octavian Brabete},
  \bibinfo{person}{Dmitrii Ustiugov}, \bibinfo{person}{Yuvraj Patel}, {and}
  \bibinfo{person}{Luo Mai}.} \bibinfo{year}{2024}\natexlab{}.
\newblock \showarticletitle{{ServerlessLLM}: Low-Latency Serverless Inference
  for Large Language Models}. In \bibinfo{booktitle}{\emph{OSDI'24}}.
\newblock


\bibitem[Gim et~al\mbox{.}(2024)]%
        {promptcache}
\bibfield{author}{\bibinfo{person}{In Gim}, \bibinfo{person}{Guojun Chen},
  \bibinfo{person}{Seung seob Lee}, \bibinfo{person}{Nikhil Sarda},
  \bibinfo{person}{Anurag Khandelwal}, {and} \bibinfo{person}{Lin Zhong}.}
  \bibinfo{year}{2024}\natexlab{}.
\newblock \bibinfo{title}{Prompt Cache: Modular Attention Reuse for Low-Latency
  Inference}.
\newblock
\showeprint[arxiv]{2311.04934}~[cs.CL]
\urldef\tempurl%
\url{https://arxiv.org/abs/2311.04934}
\showURL{%
\tempurl}


\bibitem[{GLM-5-Team} et~al\mbox{.}(2026)]%
        {glm5}
\bibfield{author}{\bibinfo{person}{{GLM-5-Team}} {et~al\mbox{.}}}
  \bibinfo{year}{2026}\natexlab{}.
\newblock \bibinfo{title}{{GLM-5}: from Vibe Coding to Agentic Engineering}.
\newblock
\showeprint[arxiv]{2602.15763}~[cs.LG]
\urldef\tempurl%
\url{https://arxiv.org/abs/2602.15763}
\showURL{%
\tempurl}


\bibitem[Goldman et~al\mbox{.}(2026)]%
        {nccl_ep}
\bibfield{author}{\bibinfo{person}{Amos Goldman}, \bibinfo{person}{Nimrod
  Boker}, \bibinfo{person}{Maayan Sheraizin}, \bibinfo{person}{Nimrod Admoni},
  \bibinfo{person}{Artem Polyakov}, \bibinfo{person}{Subhadeep Bhattacharya},
  \bibinfo{person}{Fan Yu}, \bibinfo{person}{Kai Sun},
  \bibinfo{person}{Georgios Theodorakis}, \bibinfo{person}{Hsin-Chun Yin},
  \bibinfo{person}{Peter-Jan Gootzen}, \bibinfo{person}{Aamir Shafi},
  \bibinfo{person}{Assaf Ravid}, \bibinfo{person}{Salvatore~Di Girolamo},
  \bibinfo{person}{James Dinan}, \bibinfo{person}{Xiaofan Li},
  \bibinfo{person}{Manjunath~Gorentla Venkata}, {and} \bibinfo{person}{Gil
  Bloch}.} \bibinfo{year}{2026}\natexlab{}.
\newblock \bibinfo{title}{{NCCL EP}: Towards a Unified Expert Parallel
  Communication {API} for {NCCL}}.
\newblock
\showeprint[arxiv]{2603.13606}~[cs.DC]
\urldef\tempurl%
\url{https://arxiv.org/abs/2603.13606}
\showURL{%
\tempurl}


\bibitem[Hamidouche et~al\mbox{.}(2025)]%
        {ncclgin}
\bibfield{author}{\bibinfo{person}{Khaled Hamidouche}, \bibinfo{person}{John
  Bachan}, \bibinfo{person}{Pak Markthub}, \bibinfo{person}{Peter-Jan Gootzen},
  \bibinfo{person}{Elena Agostini}, \bibinfo{person}{Sylvain Jeaugey},
  \bibinfo{person}{Aamir Shafi}, \bibinfo{person}{Georgios Theodorakis}, {and}
  \bibinfo{person}{Manjunath~Gorentla Venkata}.}
  \bibinfo{year}{2025}\natexlab{}.
\newblock \bibinfo{title}{GPU-Initiated Networking for {NCCL}}.
\newblock
\showeprint[arxiv]{2511.15076}~[cs.DC]
\urldef\tempurl%
\url{https://arxiv.org/abs/2511.15076}
\showURL{%
\tempurl}


\bibitem[He et~al\mbox{.}(2026)]%
        {banaserve}
\bibfield{author}{\bibinfo{person}{Yiyuan He}, \bibinfo{person}{Minxian Xu},
  \bibinfo{person}{Jingfeng Wu}, \bibinfo{person}{Jianmin Hu},
  \bibinfo{person}{Chong Ma}, \bibinfo{person}{Min Shen}, \bibinfo{person}{Le
  Chen}, \bibinfo{person}{Chengzhong Xu}, \bibinfo{person}{Lin Qu}, {and}
  \bibinfo{person}{Kejiang Ye}.} \bibinfo{year}{2026}\natexlab{}.
\newblock \showarticletitle{BanaServe: Unified {KV Cache} and Dynamic Module
  Migration for Balancing Disaggregated {LLM} Serving in AI Infrastructure}.
\newblock \bibinfo{journal}{\emph{Software: Practice and Experience}}
  \bibinfo{volume}{56}, \bibinfo{number}{4} (\bibinfo{year}{2026}),
  \bibinfo{pages}{424--444}.
\newblock
\showeprint{https://onlinelibrary.wiley.com/doi/pdf/10.1002/spe.70054}
\href{https://doi.org/10.1002/spe.70054}{doi:\nolinkurl{10.1002/spe.70054}}


\bibitem[Hu et~al\mbox{.}(2025)]%
        {epic}
\bibfield{author}{\bibinfo{person}{Junhao Hu}, \bibinfo{person}{Wenrui Huang},
  \bibinfo{person}{Weidong Wang}, \bibinfo{person}{Haoyi Wang},
  \bibinfo{person}{Tiancheng Hu}, \bibinfo{person}{Qin Zhang},
  \bibinfo{person}{Hao Feng}, \bibinfo{person}{Xusheng Chen},
  \bibinfo{person}{Yizhou Shan}, {and} \bibinfo{person}{Tao Xie}.}
  \bibinfo{year}{2025}\natexlab{}.
\newblock \showarticletitle{{EPIC}: efficient position-independent caching for
  serving large language models}. In \bibinfo{booktitle}{\emph{Proceedings of
  the 42nd International Conference on Machine Learning}} (Vancouver, Canada)
  \emph{(\bibinfo{series}{ICML'25})}. \bibinfo{publisher}{JMLR.org}, Article
  \bibinfo{articleno}{956}, \bibinfo{numpages}{12}~pages.
\newblock


\bibitem[Jiang et~al\mbox{.}(2026a)]%
        {moecap}
\bibfield{author}{\bibinfo{person}{Yinsicheng Jiang}, \bibinfo{person}{Yao Fu},
  \bibinfo{person}{Yeqi Huang}, \bibinfo{person}{Ping Nie},
  \bibinfo{person}{Zhan Lu}, \bibinfo{person}{Leyang Xue},
  \bibinfo{person}{Congjie He}, \bibinfo{person}{Man-Kit Sit},
  \bibinfo{person}{Jilong Xue}, \bibinfo{person}{Li Dong},
  \bibinfo{person}{Ziming Miao}, \bibinfo{person}{DaYou Du},
  \bibinfo{person}{Tairan Xu}, \bibinfo{person}{Kai Zou},
  \bibinfo{person}{Edoardo Ponti}, {and} \bibinfo{person}{Luo Mai}.}
  \bibinfo{year}{2026}\natexlab{a}.
\newblock \showarticletitle{MoE-{CAP}: Benchmarking Cost, Accuracy and
  Performance of Sparse Mixture-of-Experts Systems}. In
  \bibinfo{booktitle}{\emph{The Thirty-ninth Annual Conference on Neural
  Information Processing Systems Datasets and Benchmarks Track}}.
\newblock
\urldef\tempurl%
\url{https://openreview.net/forum?id=k2fWVhG0u5}
\showURL{%
\tempurl}


\bibitem[Jiang et~al\mbox{.}(2026b)]%
        {contextpilot}
\bibfield{author}{\bibinfo{person}{Yinsicheng Jiang}, \bibinfo{person}{Yeqi
  Huang}, \bibinfo{person}{Liang Cheng}, \bibinfo{person}{Cheng Deng},
  \bibinfo{person}{Xuan Sun}, {and} \bibinfo{person}{Luo Mai}.}
  \bibinfo{year}{2026}\natexlab{b}.
\newblock \showarticletitle{{ContextPilot}: Fast Long-Context Inference via
  Context Reuse}. In \bibinfo{booktitle}{\emph{Proceedings of the 9th
  Conference on Machine Learning and Systems (MLSys 2026)}}.
\newblock
\urldef\tempurl%
\url{https://arxiv.org/abs/2511.03475}
\showURL{%
\tempurl}


\bibitem[Jiashi~Li(2025)]%
        {flashmla}
\bibfield{author}{\bibinfo{person}{Shengyu~Liu Jiashi~Li}.}
  \bibinfo{year}{2025}\natexlab{}.
\newblock \bibinfo{title}{FlashMLA: Efficient Multi-head Latent Attention
  Kernels}.
\newblock
  \bibinfo{howpublished}{\url{https://github.com/deepseek-ai/FlashMLA}}.
\newblock


\bibitem[{Kimi Team} et~al\mbox{.}(2026)]%
        {kimik2}
\bibfield{author}{\bibinfo{person}{{Kimi Team}} {et~al\mbox{.}}}
  \bibinfo{year}{2026}\natexlab{}.
\newblock \bibinfo{title}{Kimi K2: Open Agentic Intelligence}.
\newblock
\showeprint[arxiv]{2507.20534}~[cs.LG]
\urldef\tempurl%
\url{https://arxiv.org/abs/2507.20534}
\showURL{%
\tempurl}


\bibitem[Kwon et~al\mbox{.}(2023)]%
        {vllm}
\bibfield{author}{\bibinfo{person}{Woosuk Kwon}, \bibinfo{person}{Zhuohan Li},
  \bibinfo{person}{Siyuan Zhuang}, \bibinfo{person}{Ying Sheng},
  \bibinfo{person}{Lianmin Zheng}, \bibinfo{person}{Cody~Hao Yu},
  \bibinfo{person}{Joseph Gonzalez}, \bibinfo{person}{Hao Zhang}, {and}
  \bibinfo{person}{Ion Stoica}.} \bibinfo{year}{2023}\natexlab{}.
\newblock \showarticletitle{Efficient Memory Management for Large Language
  Model Serving with PagedAttention}. In \bibinfo{booktitle}{\emph{Proceedings
  of the 29th Symposium on Operating Systems Principles}} (Koblenz, Germany)
  \emph{(\bibinfo{series}{SOSP '23})}. \bibinfo{publisher}{Association for
  Computing Machinery}, \bibinfo{address}{New York, NY, USA},
  \bibinfo{pages}{611--626}.
\newblock
\showISBNx{9798400702297}
\href{https://doi.org/10.1145/3600006.3613165}{doi:\nolinkurl{10.1145/3600006.3613165}}


\bibitem[Levy(2026)]%
        {dsaccess}
\bibfield{author}{\bibinfo{person}{Noam Levy}.}
  \bibinfo{year}{2026}\natexlab{}.
\newblock \bibinfo{title}{Dynamic Sparse Attention: Access Patterns and
  Architecture}.
\newblock
\showeprint[arxiv]{2603.13430}~[cs.AR]
\urldef\tempurl%
\url{https://arxiv.org/abs/2603.13430}
\showURL{%
\tempurl}


\bibitem[Liang et~al\mbox{.}(2025)]%
        {adrenaline}
\bibfield{author}{\bibinfo{person}{Yunkai Liang}, \bibinfo{person}{Zhangyu
  Chen}, \bibinfo{person}{Pengfei Zuo}, \bibinfo{person}{Zhi Zhou},
  \bibinfo{person}{Xu Chen}, {and} \bibinfo{person}{Zhou Yu}.}
  \bibinfo{year}{2025}\natexlab{}.
\newblock \bibinfo{title}{Injecting Adrenaline into {LLM} Serving: Boosting
  Resource Utilization and Throughput via Attention Disaggregation}.
\newblock
\showeprint[arxiv]{2503.20552}~[cs.DC]
\urldef\tempurl%
\url{https://arxiv.org/abs/2503.20552}
\showURL{%
\tempurl}


\bibitem[Licker et~al\mbox{.}(2026)]%
        {transferengine}
\bibfield{author}{\bibinfo{person}{Nandor Licker}, \bibinfo{person}{Kevin Hu},
  \bibinfo{person}{Vladimir Zaytsev}, {and} \bibinfo{person}{Lequn Chen}.}
  \bibinfo{year}{2026}\natexlab{}.
\newblock \bibinfo{title}{fabric-lib: RDMA Point-to-Point Communication for
  {LLM} Systems}.
\newblock
\showeprint[arxiv]{2510.27656}~[cs.DC]
\urldef\tempurl%
\url{https://arxiv.org/abs/2510.27656}
\showURL{%
\tempurl}


\bibitem[Lin et~al\mbox{.}(2024)]%
        {distattention}
\bibfield{author}{\bibinfo{person}{Bin Lin}, \bibinfo{person}{Chen Zhang},
  \bibinfo{person}{Tao Peng}, \bibinfo{person}{Hanyu Zhao},
  \bibinfo{person}{Wencong Xiao}, \bibinfo{person}{Minmin Sun},
  \bibinfo{person}{Anmin Liu}, \bibinfo{person}{Zhipeng Zhang},
  \bibinfo{person}{Lanbo Li}, \bibinfo{person}{Xiafei Qiu},
  \bibinfo{person}{Shen Li}, \bibinfo{person}{Zhigang Ji}, \bibinfo{person}{Tao
  Xie}, \bibinfo{person}{Yong Li}, {and} \bibinfo{person}{Wei Lin}.}
  \bibinfo{year}{2024}\natexlab{}.
\newblock \bibinfo{title}{Infinite-LLM: Efficient {LLM} Service for Long
  Context with DistAttention and Distributed KVCache}.
\newblock
\showeprint[arxiv]{2401.02669}~[cs.DC]
\urldef\tempurl%
\url{https://arxiv.org/abs/2401.02669}
\showURL{%
\tempurl}


\bibitem[Mao et~al\mbox{.}(2025)]%
        {uccl_ep}
\bibfield{author}{\bibinfo{person}{Ziming Mao}, \bibinfo{person}{Yihan Zhang},
  \bibinfo{person}{Chihan Cui}, \bibinfo{person}{Zhen Huang},
  \bibinfo{person}{Kaichao You}, \bibinfo{person}{Zhongjie Chen},
  \bibinfo{person}{Zhiying Xu}, \bibinfo{person}{Zhenyu Gu},
  \bibinfo{person}{Scott Shenker}, \bibinfo{person}{Costin Raiciu},
  \bibinfo{person}{Yang Zhou}, {and} \bibinfo{person}{Ion Stoica}.}
  \bibinfo{year}{2025}\natexlab{}.
\newblock \bibinfo{title}{{UCCL-EP}: Portable Expert-Parallel Communication}.
\newblock
\showeprint[arxiv]{2512.19849}~[cs.DC]
\urldef\tempurl%
\url{https://arxiv.org/abs/2512.19849}
\showURL{%
\tempurl}


\bibitem[Milakov and Gimelshein(2018)]%
        {onlinesoftmax}
\bibfield{author}{\bibinfo{person}{Maxim Milakov} {and}
  \bibinfo{person}{Natalia Gimelshein}.} \bibinfo{year}{2018}\natexlab{}.
\newblock \showarticletitle{Online normalizer calculation for softmax}.
\newblock \bibinfo{journal}{\emph{CoRR}}  \bibinfo{volume}{abs/1805.02867}
  (\bibinfo{year}{2018}).
\newblock
\showeprint[arXiv]{1805.02867}
\urldef\tempurl%
\url{http://arxiv.org/abs/1805.02867}
\showURL{%
\tempurl}


\bibitem[{Moonshot AI}(2026)]%
        {kimik26}
\bibfield{author}{\bibinfo{person}{{Moonshot AI}}.}
  \bibinfo{year}{2026}\natexlab{}.
\newblock \bibinfo{title}{{Kimi K2.6}}.
\newblock \bibinfo{howpublished}{Hugging Face model card}.
\newblock
\urldef\tempurl%
\url{https://huggingface.co/moonshotai/Kimi-K2.6}
\showURL{%
\tempurl}


\bibitem[Qin et~al\mbox{.}(2025)]%
        {mooncake}
\bibfield{author}{\bibinfo{person}{Ruoyu Qin}, \bibinfo{person}{Zheming Li},
  \bibinfo{person}{Weiran He}, \bibinfo{person}{Jialei Cui},
  \bibinfo{person}{Heyi Tang}, \bibinfo{person}{Feng Ren},
  \bibinfo{person}{Teng Ma}, \bibinfo{person}{Shangming Cai},
  \bibinfo{person}{Yineng Zhang}, \bibinfo{person}{Mingxing Zhang},
  \bibinfo{person}{Yongwei Wu}, \bibinfo{person}{Weimin Zheng}, {and}
  \bibinfo{person}{Xinran Xu}.} \bibinfo{year}{2025}\natexlab{}.
\newblock \showarticletitle{Mooncake: A KVCache-centric Disaggregated
  Architecture for {LLM} Serving}.
\newblock \bibinfo{journal}{\emph{ACM Trans. Storage}} (\bibinfo{date}{Nov.}
  \bibinfo{year}{2025}).
\newblock
\showISSN{1553-3077}
\href{https://doi.org/10.1145/3773772}{doi:\nolinkurl{10.1145/3773772}}
\newblock
\shownote{Just Accepted}.


\bibitem[Takbir et~al\mbox{.}(2025)]%
        {flexicache}
\bibfield{author}{\bibinfo{person}{Nazmul Takbir}, \bibinfo{person}{Hamidreza
  Alikhani}, \bibinfo{person}{Nikil Dutt}, {and}
  \bibinfo{person}{Sangeetha~Abdu Jyothi}.} \bibinfo{year}{2025}\natexlab{}.
\newblock \bibinfo{title}{{FlexiCache}: Leveraging Temporal Stability of
  Attention Heads for Efficient {KV} Cache Management}.
\newblock
\showeprint[arxiv]{2511.00868}~[cs.LG]
\urldef\tempurl%
\url{https://arxiv.org/abs/2511.00868}
\showURL{%
\tempurl}


\bibitem[Tang et~al\mbox{.}(2026)]%
        {tpla}
\bibfield{author}{\bibinfo{person}{Xiaojuan Tang}, \bibinfo{person}{Fanxu
  Meng}, \bibinfo{person}{Pingzhi Tang}, \bibinfo{person}{Yuxuan Wang},
  \bibinfo{person}{Di Yin}, \bibinfo{person}{Xing Sun}, {and}
  \bibinfo{person}{Muhan Zhang}.} \bibinfo{year}{2026}\natexlab{}.
\newblock \showarticletitle{{TPLA}: Tensor Parallel Latent Attention for
  Efficient Disaggregated Prefill \& Decode Inference}. In
  \bibinfo{booktitle}{\emph{Proceedings of the 31st ACM International
  Conference on Architectural Support for Programming Languages and Operating
  Systems, Volume 2}} (USA) \emph{(\bibinfo{series}{ASPLOS '26})}.
  \bibinfo{publisher}{Association for Computing Machinery},
  \bibinfo{address}{New York, NY, USA}, \bibinfo{pages}{2048--2062}.
\newblock
\showISBNx{9798400723599}
\href{https://doi.org/10.1145/3779212.3790237}{doi:\nolinkurl{10.1145/3779212.3790237}}


\bibitem[Wang et~al\mbox{.}(2025)]%
        {mepic}
\bibfield{author}{\bibinfo{person}{Qian Wang}, \bibinfo{person}{Zahra
  Yousefijamarani}, \bibinfo{person}{Morgan~Lindsay Heisler},
  \bibinfo{person}{Rongzhi Gu}, \bibinfo{person}{Bai Xiaolong},
  \bibinfo{person}{Shan Yizhou}, \bibinfo{person}{Wei Zhang},
  \bibinfo{person}{Wang Lan}, \bibinfo{person}{Ying Xiong},
  \bibinfo{person}{Yong Zhang}, {and} \bibinfo{person}{Zhenan Fan}.}
  \bibinfo{year}{2025}\natexlab{}.
\newblock \bibinfo{title}{{MEPIC}: Memory Efficient Position Independent
  Caching for {LLM} Serving}.
\newblock
\showeprint[arxiv]{2512.16822}~[cs.LG]
\urldef\tempurl%
\url{https://arxiv.org/abs/2512.16822}
\showURL{%
\tempurl}


\bibitem[Williams et~al\mbox{.}(2009)]%
        {roofline}
\bibfield{author}{\bibinfo{person}{Samuel Williams}, \bibinfo{person}{Andrew
  Waterman}, {and} \bibinfo{person}{David Patterson}.}
  \bibinfo{year}{2009}\natexlab{}.
\newblock \showarticletitle{Roofline: an insightful visual performance model
  for multicore architectures}.
\newblock \bibinfo{journal}{\emph{Commun. ACM}} \bibinfo{volume}{52},
  \bibinfo{number}{4} (\bibinfo{date}{April} \bibinfo{year}{2009}),
  \bibinfo{pages}{65--76}.
\newblock
\showISSN{0001-0782}
\href{https://doi.org/10.1145/1498765.1498785}{doi:\nolinkurl{10.1145/1498765.1498785}}


\bibitem[Yang et~al\mbox{.}(2025)]%
        {cp_meta}
\bibfield{author}{\bibinfo{person}{Amy Yang}, \bibinfo{person}{Jingyi Yang},
  \bibinfo{person}{Aya Ibrahim}, \bibinfo{person}{Xinfeng Xie},
  \bibinfo{person}{Bangsheng Tang}, \bibinfo{person}{Grigory Sizov},
  \bibinfo{person}{Jeremy Reizenstein}, \bibinfo{person}{Jongsoo Park}, {and}
  \bibinfo{person}{Jianyu Huang}.} \bibinfo{year}{2025}\natexlab{}.
\newblock \bibinfo{title}{Context Parallelism for Scalable Million-Token
  Inference}.
\newblock
\showeprint[arxiv]{2411.01783}~[cs.DC]
\urldef\tempurl%
\url{https://arxiv.org/abs/2411.01783}
\showURL{%
\tempurl}


\bibitem[Yao et~al\mbox{.}(2026)]%
        {fluxion}
\bibfield{author}{\bibinfo{person}{Feiyu Yao}, \bibinfo{person}{Zhixiong Niu},
  \bibinfo{person}{Xiaqing Li}, \bibinfo{person}{Yongqiang Xiong},
  \bibinfo{person}{Juan Fang}, {and} \bibinfo{person}{Qian Wang}.}
  \bibinfo{year}{2026}\natexlab{}.
\newblock \bibinfo{title}{An Efficient Hybrid Sparse Attention with {CPU}-{GPU}
  Parallelism for Long-Context Inference}.
\newblock
\showeprint[arxiv]{2605.07719}~[cs.LG]
\urldef\tempurl%
\url{https://arxiv.org/abs/2605.07719}
\showURL{%
\tempurl}


\bibitem[Yao et~al\mbox{.}(2025)]%
        {cacheblend}
\bibfield{author}{\bibinfo{person}{Jiayi Yao}, \bibinfo{person}{Hanchen Li},
  \bibinfo{person}{Yuhan Liu}, \bibinfo{person}{Siddhant Ray},
  \bibinfo{person}{Yihua Cheng}, \bibinfo{person}{Qizheng Zhang},
  \bibinfo{person}{Kuntai Du}, \bibinfo{person}{Shan Lu}, {and}
  \bibinfo{person}{Junchen Jiang}.} \bibinfo{year}{2025}\natexlab{}.
\newblock \showarticletitle{CacheBlend: Fast Large Language Model Serving for
  RAG with Cached Knowledge Fusion}. In \bibinfo{booktitle}{\emph{Proceedings
  of the Twentieth European Conference on Computer Systems}}.
  \bibinfo{pages}{94--109}.
\newblock
\href{https://doi.org/10.1145/3689031.3696098}{doi:\nolinkurl{10.1145/3689031.3696098}}


\bibitem[Ye et~al\mbox{.}(2025)]%
        {flashinfer}
\bibfield{author}{\bibinfo{person}{Zihao Ye}, \bibinfo{person}{Lequn Chen},
  \bibinfo{person}{Ruihang Lai}, \bibinfo{person}{Wuwei Lin},
  \bibinfo{person}{Yineng Zhang}, \bibinfo{person}{Stephanie Wang},
  \bibinfo{person}{Tianqi Chen}, \bibinfo{person}{Baris Kasikci},
  \bibinfo{person}{Vinod Grover}, \bibinfo{person}{Arvind Krishnamurthy}, {and}
  \bibinfo{person}{Luis Ceze}.} \bibinfo{year}{2025}\natexlab{}.
\newblock \showarticletitle{{FlashInfer}: Efficient and Customizable Attention
  Engine for {LLM} Inference Serving}.
\newblock \bibinfo{journal}{\emph{arXiv preprint arXiv:2501.01005}}
  (\bibinfo{year}{2025}).
\newblock
\urldef\tempurl%
\url{https://arxiv.org/abs/2501.01005}
\showURL{%
\tempurl}


\bibitem[Yuan et~al\mbox{.}(2025)]%
        {nsa}
\bibfield{author}{\bibinfo{person}{Jingyang Yuan}, \bibinfo{person}{Huazuo
  Gao}, \bibinfo{person}{Damai Dai}, \bibinfo{person}{Junyu Luo},
  \bibinfo{person}{Liang Zhao}, \bibinfo{person}{Zhengyan Zhang},
  \bibinfo{person}{Zhenda Xie}, \bibinfo{person}{Yuxing Wei},
  \bibinfo{person}{Lean Wang}, \bibinfo{person}{Zhiping Xiao},
  \bibinfo{person}{Yuqing Wang}, \bibinfo{person}{Chong Ruan},
  \bibinfo{person}{Ming Zhang}, \bibinfo{person}{Wenfeng Liang}, {and}
  \bibinfo{person}{Wangding Zeng}.} \bibinfo{year}{2025}\natexlab{}.
\newblock \showarticletitle{Native Sparse Attention: Hardware-Aligned and
  Natively Trainable Sparse Attention}. In
  \bibinfo{booktitle}{\emph{Proceedings of the 63rd Annual Meeting of the
  Association for Computational Linguistics (Volume 1: Long Papers)}},
  \bibfield{editor}{\bibinfo{person}{Wanxiang Che}, \bibinfo{person}{Joyce
  Nabende}, \bibinfo{person}{Ekaterina Shutova}, {and}
  \bibinfo{person}{Mohammad~Taher Pilehvar}} (Eds.).
  \bibinfo{publisher}{Association for Computational Linguistics},
  \bibinfo{address}{Vienna, Austria}, \bibinfo{pages}{23078--23097}.
\newblock
\showISBNx{979-8-89176-251-0}
\href{https://doi.org/10.18653/v1/2025.acl-long.1126}{doi:\nolinkurl{10.18653/v1/2025.acl-long.1126}}


\bibitem[Yun et~al\mbox{.}(2026)]%
        {newbottleneck}
\bibfield{author}{\bibinfo{person}{Sungmin Yun}, \bibinfo{person}{Seonyong
  Park}, \bibinfo{person}{Hwayong Nam}, \bibinfo{person}{Younjoo Lee},
  \bibinfo{person}{Gunjun Lee}, \bibinfo{person}{Kwanhee Kyung},
  \bibinfo{person}{Sangpyo Kim}, \bibinfo{person}{Nam~Sung Kim},
  \bibinfo{person}{Jongmin Kim}, \bibinfo{person}{Hyungyo Kim},
  \bibinfo{person}{Juhwan Cho}, \bibinfo{person}{Seungmin Baek}, {and}
  \bibinfo{person}{Jung~Ho Ahn}.} \bibinfo{year}{2026}\natexlab{}.
\newblock \bibinfo{title}{Rethinking {LLM} Inference Bottlenecks: Insights from
  Latent Attention and Mixture-of-Experts}.
\newblock
\showeprint[arxiv]{2507.15465}~[cs.AR]
\urldef\tempurl%
\url{https://arxiv.org/abs/2507.15465}
\showURL{%
\tempurl}


\bibitem[{Z.ai}(2026)]%
        {glm51}
\bibfield{author}{\bibinfo{person}{{Z.ai}}.} \bibinfo{year}{2026}\natexlab{}.
\newblock \bibinfo{title}{{GLM-5.1}}.
\newblock \bibinfo{howpublished}{Hugging Face model card}.
\newblock
\urldef\tempurl%
\url{https://huggingface.co/zai-org/GLM-5.1}
\showURL{%
\tempurl}


\bibitem[Zhang et~al\mbox{.}(2023a)]%
        {multichip}
\bibfield{author}{\bibinfo{person}{Shiqing Zhang}, \bibinfo{person}{Mahmood
  Naderan-Tahan}, \bibinfo{person}{Magnus Jahre}, {and} \bibinfo{person}{Lieven
  Eeckhout}.} \bibinfo{year}{2023}\natexlab{a}.
\newblock \showarticletitle{Characterizing Multi-Chip GPU Data Sharing}.
\newblock \bibinfo{journal}{\emph{ACM Trans. Archit. Code Optim.}}
  \bibinfo{volume}{20}, \bibinfo{number}{4}, Article \bibinfo{articleno}{56}
  (\bibinfo{date}{Dec.} \bibinfo{year}{2023}), \bibinfo{numpages}{24}~pages.
\newblock
\showISSN{1544-3566}
\href{https://doi.org/10.1145/3629521}{doi:\nolinkurl{10.1145/3629521}}


\bibitem[Zhang et~al\mbox{.}(2023b)]%
        {sac}
\bibfield{author}{\bibinfo{person}{Shiqing Zhang}, \bibinfo{person}{Mahmood
  Naderan-Tahan}, \bibinfo{person}{Magnus Jahre}, {and} \bibinfo{person}{Lieven
  Eeckhout}.} \bibinfo{year}{2023}\natexlab{b}.
\newblock \showarticletitle{SAC: Sharing-Aware Caching in Multi-Chip GPUs}. In
  \bibinfo{booktitle}{\emph{Proceedings of the 50th Annual International
  Symposium on Computer Architecture}} (Orlando, FL, USA)
  \emph{(\bibinfo{series}{ISCA '23})}. \bibinfo{publisher}{Association for
  Computing Machinery}, \bibinfo{address}{New York, NY, USA}, Article
  \bibinfo{articleno}{43}, \bibinfo{numpages}{13}~pages.
\newblock
\showISBNx{9798400700958}
\href{https://doi.org/10.1145/3579371.3589078}{doi:\nolinkurl{10.1145/3579371.3589078}}


\bibitem[Zhao et~al\mbox{.}(2025)]%
        {deepep}
\bibfield{author}{\bibinfo{person}{Chenggang Zhao}, \bibinfo{person}{Shangyan
  Zhou}, \bibinfo{person}{Liyue Zhang}, \bibinfo{person}{Chengqi Deng},
  \bibinfo{person}{Zhean Xu}, \bibinfo{person}{Yuxuan Liu},
  \bibinfo{person}{Kuai Yu}, \bibinfo{person}{Jiashi Li}, {and}
  \bibinfo{person}{Liang Zhao}.} \bibinfo{year}{2025}\natexlab{}.
\newblock \bibinfo{title}{{DeepEP}: an efficient expert-parallel communication
  library}.
\newblock \bibinfo{howpublished}{\url{https://github.com/deepseek-ai/DeepEP}}.
\newblock


\bibitem[Zheng et~al\mbox{.}(2024)]%
        {sglang}
\bibfield{author}{\bibinfo{person}{Lianmin Zheng}, \bibinfo{person}{Liangsheng
  Yin}, \bibinfo{person}{Zhiqiang Xie}, \bibinfo{person}{Chuyue Sun},
  \bibinfo{person}{Jeff Huang}, \bibinfo{person}{Cody~Hao Yu},
  \bibinfo{person}{Shiyi Cao}, \bibinfo{person}{Christos Kozyrakis},
  \bibinfo{person}{Ion Stoica}, \bibinfo{person}{Joseph~E. Gonzalez},
  \bibinfo{person}{Clark Barrett}, {and} \bibinfo{person}{Ying Sheng}.}
  \bibinfo{year}{2024}\natexlab{}.
\newblock \showarticletitle{SGLang: efficient execution of structured language
  model programs}. In \bibinfo{booktitle}{\emph{Proceedings of the 38th
  International Conference on Neural Information Processing Systems}}
  (Vancouver, BC, Canada) \emph{(\bibinfo{series}{NIPS '24})}.
  \bibinfo{publisher}{Curran Associates Inc.}, \bibinfo{address}{Red Hook, NY,
  USA}, Article \bibinfo{articleno}{2000}, \bibinfo{numpages}{27}~pages.
\newblock
\showISBNx{9798331314385}


\bibitem[Zhong et~al\mbox{.}(2024)]%
        {distserve}
\bibfield{author}{\bibinfo{person}{Yinmin Zhong}, \bibinfo{person}{Shengyu
  Liu}, \bibinfo{person}{Junda Chen}, \bibinfo{person}{Jianbo Hu},
  \bibinfo{person}{Yibo Zhu}, \bibinfo{person}{Xuanzhe Liu},
  \bibinfo{person}{Xin Jin}, {and} \bibinfo{person}{Hao Zhang}.}
  \bibinfo{year}{2024}\natexlab{}.
\newblock \showarticletitle{{DistServe}: disaggregating prefill and decoding
  for goodput-optimized large language model serving}. In
  \bibinfo{booktitle}{\emph{Proceedings of the 18th USENIX Conference on
  Operating Systems Design and Implementation}} (Santa Clara, CA, USA)
  \emph{(\bibinfo{series}{OSDI'24})}. \bibinfo{publisher}{USENIX Association},
  \bibinfo{address}{USA}, Article \bibinfo{articleno}{11},
  \bibinfo{numpages}{18}~pages.
\newblock
\showISBNx{978-1-939133-40-3}


\end{thebibliography}

\end{document}